\documentclass[10pt]{article} 
\usepackage{amsmath,amsthm,amsfonts,amssymb,epsfig} 
 
\usepackage{graphics} 
\usepackage{verbatim} 
 
\newcommand{\reals}{\mathbb{R}}

\newcommand{\interior}{\operatorname{int}} 
 
\newcommand{\diag}{\operatorname{diag}} 
 
\newcommand{\deter}{\operatorname{det}} 
\newcommand{\trace}{\operatorname{tr}}

\theoremstyle{plain} 
\newtheorem{defi}{Definition} 
\newtheorem{prop}{Proposition} 
\newtheorem{stel}{Theorem} 
 
\newtheorem{lemma}{Lemma} 
\theoremstyle{remark}

\begin{document} 
\title{Multi-strain virus dynamics with mutations:\\ A global analysis} 
 
\author{
Patrick De Leenheer\footnote{email: {\tt deleenhe@math.ufl.edu}. Supported in part by NSF grant DMS-0614651.}
and 
Sergei S. Pilyugin\footnote{email: {\tt pilyugin@math.ufl.edu}. Supported in part by NSF grant DMS-0517954.}
\\ 
Department of Mathematics\\University of Florida, Gainesville, FL 32611-8105, USA\\[3ex] 
{\it To our mentor and good friend Hal Smith, on the occasion of his $60$th birthday.}} 
 
\date{} 
\maketitle 
 
\begin{abstract} 
We consider within-host virus models with $n\geq 2$ strains and allow mutation between the strains.  
If there is no mutation, a  
Lyapunov function establishes global stability of the steady state corresponding to the fittest strain.  
For small perturbations this steady state persists, perhaps with small concentrations of some or all other strains,  
depending on the connectivity of the graph describing all possible mutations.  
Moreover, using a perturbation result due to Smith and Waltman \cite{smith-waltman}, we show that this steady state also  
preserves global stability.  
\end{abstract} 
 
\section{Introduction} 
The study of the dynamics of within-host virus disease models has been a very fruitful area of research over  
the past few decades. Of particular importance has been the work on mathematical  
models of HIV infection by Perelson and coauthors \cite{perelson-etal,perelson-nelson} and Nowak and coauthors  
\cite{nowak-may}. It has spurred more recent research by among others Hal Smith with one of us \cite{hiv},  
\cite{wang-li} and \cite{zhilan}.  
 
For single-strain virus models, the understanding of the global behavior has been largely based on  
the fact that they are competitive \cite{hiv} and the use of particular mathematical tools developed by   
Muldowney; see Li and Muldowney \cite{li-muldowney} for an application of these techniques to the   
classical SEIR model in epidemiology.  
Of course it is well known that for globally stable systems there is a Lyapunov function, but  
finding such a function is often difficult, as illustrated by the following quote from  
Smith and Waltman's classical work on chemostats \cite{smith-waltman-chem} on p. 37: 

\vskip 0.1in 
\centerline{\it Considerable ingenuity, intuition, and perhaps luck are required to find a Liapunov function.} 
\vskip 0.1in  
 
One of the purposes of this paper is to find such Lyapunov functions for various within host virus models following  
the ingenuity from \cite{korobeinikov} and \cite{iggidr}. Another purpose of the paper is to investigate  
what happens if we include mutation effects in the model  
by allowing different virus strains to mutate into each other. This is very relevant in the context of HIV where  
mutations have profound impact on treatment, see for instance \cite{zhilan} where a two-strain model is considered. 
 
Mathematically we will treat the model with mutations as a perturbation of the original model.  
It turns out that the structural properties of the mutation matrix that describes the possible mutations  
(in particular, whether this matrix is irreducible or not),  
dictate which single strain steady states of the unperturbed model persist in the perturbed model, and which don't.  
An obvious problem is to determine if the globally stable single strain steady state of the unperturbed model persists.  
We will show that this is always the case, regardless of the mutation matrix.   
Moreover, taking advantage of the perturbation result developed by Smith and Waltman  
in \cite{smith-waltman}, we will show that this steady state remains globally stable  
for small values of the mutation parameter. 
In order to apply this perturbation result we will first need to establish a particular persistence  
property, uniform in the perturbation parameter, and to achieve this we invoke the theory developed  
by Hutson \cite{hutson1,hutson2}, see also \cite{hofbauer}, which uses the notion of an average Lyapunov function.  
It will be shown that a rather simple -in fact, linear- average Lyapunov function exists. 
 
The paper is organized as follows. In Section \ref{single} we present a Lyapunov function  
to establish global stability of the disease equilibrium of a single-strain virus model. This   
is extended in Section \ref{multi} to a global stability result for a multi-strain model which 
does not include mutations. In biological terms, we demonstrate that in the absence of mutations 
the fittest strain of the virus drives all  other  viral strains to extinction.
In Section \ref{mutation} we investigate what happens if mutations are taken into account for  
two different models.  
Finally, in the Appendix we extend all our results to a slightly modified model which includes an often  
neglected loss term in the virus equation.

\section{Single-strain}\label{single} 
In this paper, we consider the basic model of the form
\begin{eqnarray}\label{hiv1} 
{\dot T}&=&f(T)-kVT\nonumber \\ 
{\dot T^*}&=&kVT-\beta T^*\nonumber \\ 
{\dot V}&=&N\beta T^*-\gamma V, 
\end{eqnarray} 
where $T$, $T^*$, $V$ denote the concentrations of uninfected (healthy) and infected host cells, and free
virions, respectively. Equations (\ref{hiv1}) describe a general viral infection
where the viral replication is limited by the availability of target cells $T$. In this
model, we assume that all infected cells $T^*$ are virus-producing cells, that is, we do
not include any intermediate stage(s) corresponding to latently infected cells. In 
addition, we do not explicitly consider the impact of the immune response. Implicitly,
the immune response can be accounted for by the removal term $-\beta T^*$. The rate of
viral production is assumed proportional to the removal of infected cells. In case
of lytic viruses, $N$ represents the average burst size of a single infected cell;
whereas in case of budding viruses, $N$ can be thought of as the average number of
virions produced over a lifetime of an infected cell. For different infections, the actual
class of the target cells in (\ref{hiv1}) may vary from the $CD4+$ $T$ lymphocytes (in case of HIV),
to the epithelial cells (in case of Influenza), to the red blood cells (in case of Malaria). The
$T$, $T^*$, $V$ notation is adopted from the classical HIV model \cite{perelson-nelson}.
  
All parameters are assumed to be positive.  
The parameters $\beta$ and $\gamma$ are the removal rates of the infected cells and virus particles 
respectively.  
Following \cite{perelson-nelson,nowak-may}, we neglect the term  in the $V$-equation that represents the  
loss of a virus particle upon infection. But all subsequent results hold when  
this loss term is included, in which case the $V$-equation reads: 
$$ 
{\dot V}=N\beta T^*-\gamma V-kVT. 
$$ 
These results will be presented in the Appendix. 
 
The growth rate of the uninfected cell population is given by the smooth function  
$f(T):\reals_+ \rightarrow \reals$, which is assumed to satisfy the following: 
\begin{equation}\label{T0} 
\exists \; T_0>0 \;\; :\;\; f(T)(T-T_0)<0,\;\; T\neq T_0. 
\end{equation} 
Since continuity of $f$ implies that $f(T_0)=0$, it is easy to see that 
$$ 
E_0=(T_0,0,0), 
$$ 
is an equilibrium of $(\ref{hiv1})$. Effectively, $T_0$ is the carrying capacity for the healthy
cell population.  

A second, positive equilibrium may exist if the following quantities are positive: 
\begin{equation}\label{eq} 
{\bar T}=\frac{\gamma}{kN},\;\; {\bar T^*}=\frac{f({\bar T})}{\beta},\;\; {\bar V}=\frac{f({\bar T})}{k{\bar T}}. 
\end{equation} 
Note that this is the case if and only if $f(\frac{\gamma}{kN})>0$, or equivalently by $(\ref{T0})$  
that ${\bar T}=\frac{\gamma}{kN}<T_0$. In terms of the basic reproduction number 
$$ 
{\cal R}^0:=\frac{kN}{\gamma}T_0=\frac{T_0}{\bar T_0}, 
$$ 
existence of a positive equilibrium is therefore equivalent to 
${\cal R}^0>1.$ 
We assume henceforth that ${\cal R}^0>1$ and denote the disease steady state by $E=({\bar T},{\bar T^*},{\bar V})$.  
Let us introduce the following sector condition: 
$$ 
{\bf (C)}\;\; (f(T)-f({\bar T}))\left(1-\frac{{\bar T}}{T}\right)\leq 0. 
$$ 
Note that this condition is satisfied when $f(T)$ is a decreasing function, independently of the value of ${\bar T}$.  
For instance, \cite{nowak-may} considers  
$f(T)=c_1-c_2T$, where $c_i$ are positive constants.  
Another example \cite{perelson-nelson} is $f(T)=s+rT(1-\frac{T}{K})$ provided that $f(0)=s\geq f({\bar T})$. 
\begin{stel}\label{1strain} 
Let {\bf (C)} hold.   
Then the equilibrium $E$ is globally asymptotically stable for $(\ref{hiv1})$  
with respect to initial conditions satisfying $T^*(0)+V(0)>0$. 
\end{stel} 
\begin{proof} 
Consider the following function on $\interior(\reals^3_+)$: 
$$ 
W=\int_{{\bar T}}^T \left(1-\frac{{\bar T}}{\tau}\right) d \tau +\int_{{\bar T^*}}^{T^*} \left(1-\frac{{\bar T^*}}{\tau}\right)d \tau+ 
\frac{\beta}{N\beta}\int_{{\bar V}}^V \left(1-\frac{{\bar V}}{\tau}\right) d \tau. 
$$ 
Then  
\begin{eqnarray*} 
{\dot W}&=&(f(T)-kVT)\left(1-\frac{{\bar T}}{T} \right)+(kVT-\beta T^*)\left(1-\frac{{\bar T^*}}{T^*} \right)+ 
\frac{1}{N}(N\beta T^*-\gamma V)\left(1-\frac{{\bar V}}{V} \right)\\ 
&=&f(T)\left(1-\frac{{\bar T}}{T} \right)+kV{\bar T}-kVT\frac{{\bar T^*}}{T^*}+\beta {\bar T^*}-\beta T^*\frac{{\bar V}}{V}  
-\frac{\gamma}{N}V+\frac{\gamma}{N}{\bar V}\\ 
\end{eqnarray*} 
Since from $(\ref{eq})$ we have that $\beta{\bar T^*}=k{\bar V}{\bar T}=\frac{\gamma}{N}{\bar V}$, it follows that 
\begin{eqnarray*} 
{\dot W}&=&f(T)\left(1-\frac{{\bar T}}{T} \right)+\beta {\bar T^*}\frac{V}{{\bar V}} 
-\beta {\bar T^*}\frac{{\bar T^*}VT}{T^*{\bar V}{\bar T}}+\beta {\bar T^*}-\beta {\bar T^*}\frac{{\bar V}T^*}{V{\bar T^*}} 
-\beta T^*\frac{V}{{\bar V}}+\beta T^*\\ 
&=&(f(T)-f({\bar T}))\left(1-\frac{{\bar T}}{T} \right)+\beta T^*\left(1-\frac{{\bar T}}{T} \right) 
+\beta {\bar T^*}\frac{V}{{\bar V}} 
-\beta {\bar T^*}\frac{{\bar T^*}VT}{T^*{\bar V}{\bar T}}+\beta {\bar T^*}-\beta {\bar T^*}\frac{{\bar V}T^*}{V{\bar T^*}} 
-\beta T^*\frac{V}{{\bar V}}+\beta T^*\\ 
&=&(f(T)-f({\bar T}))\left(1-\frac{{\bar T}}{T} \right)- 
\beta {\bar T^*}\left[\frac{{\bar T}}{T}+\frac{{\bar T^*}VT}{T^*{\bar V}{\bar T}}+\frac{{\bar V}T^*}{V{\bar T^*}}-3\right] 
\end{eqnarray*} 
The first term is non-positive by {\bf (C)}.  
The second term is non-positive as well since the geometric mean of $3$  
non-negative numbers is not larger than the arithmetic mean of those numbers.  
Hence, ${\dot W}\leq 0$ in $\interior(\reals^3_+)$, and the local stability of $E$ follows.  
Notice that ${\dot W}$ equals zero iff both the first term and the second term are zero,  
and using ${\bf (C)}$, this happens at points where: 
$$ 
\frac{{\bar T}}{T}=1\textrm{ and } \frac{{\bar T^*}V}{T^*{\bar V}}=1. 
$$ 
Then LaSalle's Invariance Principle \cite{lasalle} implies that all bounded solutions  
in $\interior(\reals^3_+)$ converge to the largest invariant set in  
$$ 
M=\{(T,T^*,V)\in \interior(\reals^3_+)\;|\; \frac{{\bar T}}{T}=1,\;\; \frac{{\bar T^*}V}{T^*{\bar V}}=1\}. 
$$ 
Firstly, boundedness of all solutions follows from Lemma $\ref{bounded}$ which is proved later
in a more general setting.  
Secondly, it is clear that the largest invariant set in $M$ is the singleton $\{E\}$.  
Finally, note that forward solutions starting on  
the boundary of $\reals^3_+$ with either $T_1(0)$ or $V_1(0)$ positive,  
enter $\interior(\reals^3_+)$ instantaneously. This concludes the proof. 
\end{proof}

\section{Competitive exclusion in a multi-strain model}\label{multi} 
Let us now consider a multi-strain model: 
\begin{eqnarray} 
{\dot T}&=&f(T)-\sum_{i=1}^nk_iV_iT\label{multi1}\\ 
{\dot T^*_i}&=&k_iV_iT-\beta_iT_i^*,\;\; i=1,\dots,n\label{multi2}\\ 
{\dot V_i}&=&N_i\beta_iT_i^*-\gamma_iV_i,\;\; i=1,\dots,n\label{multi3} 
\end{eqnarray} 
where all parameters are positive.  
Similar calculations as in the single-strain model show there is a unique disease-free equilibrium  
$E_0=(T_0,0,0)$. For each $i$, there is a corresponding single-strain equilibrium $E_i$ with  
positive $T$, $T^*_i$ and $V_i$ components and zero  
components otherwise if and only if  
$$ 
1<{\cal R}^0_i. 
$$ 
Here, ${\cal R}^0_i$ is the basic reproduction number for strain $i$ which is defined by 
$$ 
{\cal R}^0_i=\frac{T_0}{{\bar T}^i}. 
$$ 
The positive components of $E_i$ are then given by  
\begin{equation}\label{poscomp} 
{\bar T}^i=\frac{\gamma_i}{k_iN_i},\;\; {\bar T}^*_i=\frac{f({\bar T}^i)}{\beta_i},\;\;  
{\bar V}_i=\frac{f({\bar T}^i)}{k_i{\bar T}^i}. 
\end{equation} 
We assume that all $E_i$ exist and assume without loss of generality (by possibly reordering components) that  
\begin{equation}\label{T's} 
{\bar T}^1<{\bar T}^2\leq \dots\leq {\bar T}^{n-1}\leq {\bar T}^n <T_0, 
\end{equation} 
or equivalently, that  
\begin{equation}\label{R's} 
1<{\cal R}^0_n\leq {\cal R}^0_{n-1}\leq \dots\leq {\cal R}^0_2<{\cal R}^0_1. 
\end{equation} 
and will prove the following competitive exclusion principle. It asserts that the  
strain with the lowest target cell concentration at steady state  
(or equivalently, with highest basic reproduction number) will ultimately dominate,  
provided that such strain is present initially. 
 
\begin{stel}\label{multistrain}   
Assume that all $E_i$ exist for $(\ref{multi1})-(\ref{multi3})$, that  
${\bf (C)}$ holds with ${\bar T}^1$ instead of ${\bar T}$, and that $(\ref{T's})$ holds.  
Then $E_1$ is globally asymptotically stable for $(\ref{multi1})-(\ref{multi3})$  
with respect to initial conditions satisfying $T_1^*(0)+V_1(0)>0$. 
\end{stel} 
\begin{proof} 
Consider the following function on $U:=\{(T,T^*_1,\dots,T^*_n,V_1,\dots,V_n)\in \reals^{2n+1}\;|\; T,T_1^*,V_1>0\}$: 
$$ 
W=\int_{{\bar T}^1}^T \left(1-\frac{{\bar T}^1}{\tau}\right) d \tau +\int_{{\bar T^*}_1}^{T^*_1} \left(1-\frac{{\bar T^*}_1}{\tau}\right)d \tau+ 
\frac{1}{N_1}\int_{{\bar V}_1}^{V_1} \left(1-\frac{{\bar V}_1}{\tau}\right) d \tau+\sum_{i=2}^n 
\left(T^*_i+\frac{1}{N_i}V_i\right). 
$$ 
Then 
\begin{eqnarray*} 
{\dot W}&=&(f(T)-\sum_{i=1}^nk_iV_iT)\left(1-\frac{{\bar T}^1}{T} \right)+(k_1V_1T-\beta_1 T_1^*)\left(1-\frac{{\bar T_1^*}}{T_1^*} \right)+ 
\frac{1}{N_1}(N_1\beta_1 T_1^*-\gamma_1 V_1)\left(1-\frac{{\bar V_1}}{V_1} \right)\\ 
&&+\sum_{i=2}^n \left(k_iV_iT-\frac{\gamma_i}{N_i}V_i\right)\\ 
&=&(f(T)-k_1V_1T)\left(1-\frac{{\bar T}^1}{T} \right)+(k_1V_1T-\beta_1 T_1^*)\left(1-\frac{{\bar T_1^*}}{T_1^*} \right)+ 
\frac{1}{N_1}(N_1\beta_1 T_1^*-\gamma_1 V_1)\left(1-\frac{{\bar V_1}}{V_1} \right)\\ 
&&-\sum_{i=2}^n \left(-k_iV_i {\bar T}^1+\frac{\gamma_i}{N_i}V_i\right)\\ 
\end{eqnarray*} 
Notice that the first three terms can be simplified in a way similar as in the proof of Theorem  
$\ref{1strain}$, and using the expression for ${\bar T}^i$ in $(\ref{poscomp})$, we find that 
\begin{eqnarray*} 
{\dot W}&=&(f(T)-f({\bar T}^1))\left(1-\frac{{\bar T}^1}{T} \right)- 
\beta_1 {\bar T^*}_1\left[\frac{{\bar T}^1}{T}+\frac{{\bar T^*}_1V_1T}{T^*_1{\bar V}_1{\bar T}^1}+ 
\frac{{\bar V}_1T^*_1}{V_1{\bar T^*}_1}-3\right]\\ 
&&-\sum_{i=2}^n k_iV_i({\bar T}^i-{\bar T}^1) 
\end{eqnarray*} 
Each of the first two terms is non-positive as was shown in the proof of Theorem $\ref{1strain}$. The third part is  
also non-positive by $(\ref{T's})$. Thus ${\dot W}\leq 0$, establishing already stability of $E_1$.  
An application of LaSalle's Invariance Principle shows that all bounded solutions in $U$  
(and as before, boundedness follows from Lemma $\ref{bounded}$ which is proved later)  
converge to the largest invariant set in  
$$ 
\left\{(T,T^*_1,\dots,T^*_n,V_1,\dots,V_n)\in U\;|\;\frac{{\bar T^1}}{T}=1,\;\; \frac{{\bar T^*_1}V_1}{T^*_1{\bar V_1}}=1 
,\;\;V_i=0,\;\; i>2 \right\}, 
$$ 
which is easily shown to be the singleton $\{E_1\}$. Finally, solutions on the boundary of $U$  
with $T_1^*(0)+V_1(0)>0$ enter $U$ instantaneously, which concludes the proof. 
\end{proof} 
 
\section{Perturbations by mutations}\label{mutation} 
In this section we expand model $(\ref{multi1})-(\ref{multi3})$ to account for mutations between the $n$ strains.  
In fact, we will study two different ways in which mutations occur. 
Our first extended model can be written compactly as follows: 
\begin{eqnarray} 
{\dot T}&=&f(T)-k'VT, \quad T \in \reals_+\label{multic1}\\ 
{\dot T^*}&=&P(\mu)KVT-B T^*, \quad T^* \in \reals^n_+\label{multic2}\\ 
{\dot V}&=&{\hat N}B T^*-\Gamma V, \quad V \in \reals^n_+\label{multic3}, 
\end{eqnarray} 
while the second is written as 
\begin{eqnarray} 
{\dot T}&=&f(T)-k'VT, \quad T \in \reals_+\label{multic4}\\ 
{\dot T^*}&=&KVT-B T^*, \quad T^* \in \reals^n_+\label{multic5}\\ 
{\dot V}&=&P(\mu){\hat N}B T^*-\Gamma V, \quad V \in \reals^n_+\label{multic6}, 
\end{eqnarray} 
In both models $K=\diag (k)$, $B=\diag(\beta)$, ${\hat N}=\diag(N)$ and $\Gamma=\diag(\gamma)$, and the matrix  
$P(\mu)$ with $\mu \in [0,1]$ is defined as follows: 
$$ 
P(\mu)=I+\mu Q, 
$$ 
where $Q$ is a matrix with $q_{ij}>0$ if strain $j$ can mutate to $i$ (for $i\neq j$) so that
different magnitudes of $q_{ij}$ reflect the possible differences in the specific
mutation rates. The diagonal  
entries of $Q$ are such that each column of $Q$ sums to zero. Notice that $P$ is a stochastic matrix  
(all its entries are in $[0,1]$ and all its columns sum to one) provided  
that $\mu \leq -1/q_{ii}$ for all $i$ (which is assumed henceforth), and that $P(0)=I$. 
 
\begin{lemma}\label{bounded} 
Both system $(\ref{multic1})-(\ref{multic3})$ and $(\ref{multic4})-(\ref{multic6})$ are dissipative, i.e.  
there exists a forward invariant compact set $K \subset \reals^{2n+1}_+$ such that every solution eventually 
enters $K$. 
\end{lemma} 
\begin{proof} 
From $(\ref{multic1})$ and $(\ref{multic4})$ follows that ${\dot T}\leq f(T)$, hence  
\begin{equation}\label{estimate} 
\limsup_{t\rightarrow \infty}T(t)\leq T_0,  
\end{equation} 
provided solutions to both systems are defined for all  
$t\geq 0$. To see that this is indeed the case, we  argue by contradiction and let  
$(T(t),T^*(t),V(t))$ be a solution with bounded maximal  interval of existence ${\cal I}_+:=[0,t_{\max})$. Then necessarily $T(t)\leq \max (T(0),T_0):=T_{\max}$ for all $t\in {\cal I}_+$. This implies that  
on ${\cal I}_+$, the following differential inequality holds for the solution of system  
$(\ref{multic1})-(\ref{multic3})$: 
\begin{eqnarray} 
{\dot T^*}&\leq&P(\mu)KVT_{\max}-B T^*\\ 
{\dot V}&\leq&{\hat N}B T^*-\Gamma V, 
\end{eqnarray} 
or for system $(\ref{multic4})-(\ref{multic6})$ 
\begin{eqnarray} 
{\dot T^*}&\leq&KVT_{\max}-B T^*\\ 
{\dot V}&\leq&P(\mu){\hat N}B T^*-\Gamma V, 
\end{eqnarray} 
respectively. Notice that the right hand sides in the above inequalities are cooperative and linear vector fields. By a comparison principle for such inequalities we obtain that $T(t)\leq {\tilde T}(t)$  
and $V(t)\leq {\tilde V}(t)$ (interpreted componentwise) for all $t$ in the intersection of the domains where the solutions are defined.  
Here, $({\tilde T}(t),\; {\tilde V}(t))$ is the solution to the linear system  
whose vector field appears in the right hand side of the above inequalities, hence 
these solutions are defined for all $t\geq 0$. But then $T(t)$ and $V(t)$ can be  
extended continuously to the closed interval $[0,T_{\max}]$, contradicting maximality of ${\cal I}_+$. 
 
Inequality $(\ref{estimate})$  implies that  for an arbitrary small $\epsilon>0$, there  
holds that $T(t)\leq T_0+\epsilon$ for all sufficiently large $t$. Now consider the behavior of the quantity  
$T+1'T^*$ along solutions of both system $(\ref{multic1})-(\ref{multic3})$ and  
$(\ref{multic4})-(\ref{multic6})$: 
$$ 
\frac{d}{dt} \left( T+1'T^*\right)=f(T)-1'BT^*\leq f(T)-b1'T^*, 
$$ 
where $b:=\min_i (\beta_i)$. By continuity of $f$ on the compact  
interval $[0,T_0+\epsilon]$, there exists (sufficiently large) $a>0$ such that
$$ f(T)+b T \leq a ,\;\; \textrm{ for all } T\in [0,T_0+\epsilon]. 
$$ 
Therefore, for all sufficiently large $t$, there holds that 
$$ 
\frac{d}{dt} \left( T+1'T^*\right)\leq a-bT-b1'T^*\leq a-b (T+1'T^*), 
$$ 
and hence  
$$ 
\limsup_{t\rightarrow \infty} T(t)+1'T^*(t)\leq \frac{a}{b}. 
$$ 
Finally, from $(\ref{multic3})$ and $(\ref{multic6})$ follows that  
$$ 
\limsup_{t\rightarrow \infty}V(t)\leq \frac{a}{b}\Gamma^{-1}{\hat N}B, 
$$ 
and 
$$ 
\limsup_{t\rightarrow \infty}V(t)\leq \frac{a}{b}\Gamma^{-1}P(\mu){\hat N}B, 
$$ 
respectively,  
where the $\limsup$ of a vector function is to be understood componentwise. Dissipativity  
now follows by observing that all the above bounds are independent of the initial condition.  
\end{proof}

\begin{lemma}\label{invertible} 
For $\mu =0$, let all single strain equilibria $E_1,E_2\dots,E_n$ exist  
for either $(\ref{multic1})-(\ref{multic3})$ or $(\ref{multic4})-(\ref{multic6})$, and assume that 
\begin{equation}\label{c1} 
{\bar T^1}<{\bar T^2}<\dots<{\bar T^n}<{\bar T^{n+1}}:=T_0, 
\end{equation} 
and  
\begin{equation}\label{c2} 
f'({\bar T}^j)\leq 0,\textrm{ for all }j=1,\dots, n+1. 
\end{equation} 
Then the Jacobian matrices of $(\ref{multic1})-(\ref{multic3})$ or $(\ref{multic4})-(\ref{multic6})$,  
evaluated at any of the $E_i$'s, $i=1,\dots, n+1$ (where $E_{n+1}:=E_0$) have the following properties: 
$J(E_i)$ has $i-1$ eigenvalues (counting multiplicities) in the open right half plane and  
$2(n+1)-i$ eigenvalues in the open left half plane. In particular, $J(E_1)$ is Hurwitz.  
\end{lemma} 
 
\begin{proof} 
Note that when $\mu=0$, the Jacobian matrix associated to  
both model $(\ref{multic1})-(\ref{multic3})$ and $(\ref{multic4})-(\ref{multic6})$ is the same and given by: 
$$ 
J=\begin{pmatrix} 
f'(T)-k'V&0&-k'T\\ 
KV&-B&KT\\ 
0&{\hat N}B&-\Gamma 
\end{pmatrix}. 
$$ 
 
To evaluate the  Jacobian  at any of the $E_i$'s it is more convenient to reorder the components of the state vector  
by means of the following permutations:  
\begin{enumerate} 
\item 
For $i=1,\dots,n$ we use  $(T,T^*,V)\rightarrow (T,T^*_i,V_i,T^*_1,V_1,\dots,T^*_{i-1},V_{i-1},T^*_{i+1},V_{i+1},\dots, T^*_n,V_n)$. 
\item 
For $i=n+1$ we use $(T,T^*,V)\rightarrow (T,T^*_1,V_1,T^*_2,V_2,\dots,T^*_n,V_n)$. 
\end{enumerate} 
Then the Jacobian matrices have the following structure: 
\begin{enumerate} 
\item 
For $i=1,\dots,n$, 
$$ 
J(E_i)= 
\begin{pmatrix} 
A_1^i&*&\dots&*&*&\dots&*\\ 
0&B_1^i&\dots&0&0&\dots&0\\ 
\vdots&\vdots&\ddots&\vdots&\vdots&\dots&0\\ 
0&0&\dots&B^i_{i-1}&0&\dots&0\\ 
0&0&\dots&0&B^i_{i+1}&\dots&0\\\ 
\vdots&\vdots&\dots&\vdots&\vdots&\ddots&0\\ 
0&0&\dots&0&0&\dots&B_n^i 
\end{pmatrix}, 
$$ 
where 
$$ 
A_1^i=\begin{pmatrix} 
f'({\bar T}^i)-k_i{\bar V}_i& 0& -k_i{\bar T}^i\\ 
k_i{\bar V}_i& -\beta_i & k_i {\bar T}^i\\\ 
0&N_i\beta_i&-\gamma_i 
\end{pmatrix}\textrm{ and } 
B_l^i=\begin{pmatrix}   
-\beta_l&k_l{\bar T}^i\\ 
N_l\beta_l& -\gamma_l 
\end{pmatrix},\; l\neq i, 
$$ 
and therefore the eigenvalues of $J(E_i)$ coincide with those of $A_1^i$ and $B_l^i$, $l\neq i$.  
Since $f'({\bar T}^i)\leq 0$ it follows from lemma $3.4$ in \cite{hiv} that the eigenvalues of $A_1^i$  
are in the open left half plane.  
The matrices $B_l^i$ are quasi-positive, irreducible matrices, hence by the Perron-Frobenius Theorem  
they have a simple real eigenvalue $\lambda_l^i$ with corresponding (componentwise) positive eigenvector.  
 
Notice that 
$$ 
\trace(B_l^i)<0, \textrm{ and } \deter(B_l^i)=\beta_l \gamma_l\left(1-\frac{{\bar T}^i}{{\bar T}^l} \right), 
$$ 
and thus by $(\ref{c1})$ that  
$$ 
\lambda_l^i\begin{cases} 
<0,\textrm{ for all } l>i,\\ 
>0,\textrm{ for all } l<i. 
\end{cases} 
$$ 
There are $i-1$ unstable $B$-blocks on the diagonal of $J(E_i)$, each of which contributes one  
positive eigenvalue  
to $J(E_i)$.

\item  
For $i=n+1$, 
$$ 
J(E_{n+1})= 
\begin{pmatrix} 
A_1^{n+1}&*&\dots&*\\ 
0&B_1^{n+1}&\dots&0\\ 
\vdots&\vdots&\ddots&\vdots\\ 
0&0&\dots&B_n^{n+1} 
\end{pmatrix}, 
$$ 
where 
$$ 
A_1^{n+1}=\begin{pmatrix} 
f'({\bar T}^{n+1}) 
\end{pmatrix}\textrm{ and } 
B_l^{n+1}=\begin{pmatrix}   
-\beta_l&k_l{\bar T}^{n+1}\\ 
N_l\beta_l& -\gamma_l 
\end{pmatrix},\; l=1,...,n. 
$$ 
Notice that by a similar argument as in the previous case, all $n$ $B$-blocks  
on the diagonal of $J(E_{n+1})$ are unstable with one positive and one negative eigenvalue.  
\end{enumerate}

\end{proof}

When $\mu \neq 0$, the question arises as to what happens to the equilibria $E_1,\dots,E_{n+1}$.
The previous Lemma allows us to apply the Implicit Function Theorem which for small positive $\mu$  
establishes the existence of (unique) equilibria $E_j(\mu)$ near each $E_j$. Indeed, denoting the vector field of  
either $(\ref{multic1})-(\ref{multic3})$ or $(\ref{multic4})-(\ref{multic6})$  
by $F(X,\mu)$, we have that for all $j=1,\dots, n+1$, there holds that  
$F(E_j,0)=0$, and under the conditions of the previous Lemma we also have that  
$\partial F / \partial X (E_j,0)$ is invertible.  It is clear that $E_{n+1}(\mu)=E_{n+1}(0)$
for all $\mu \geq 0$, i.e. the disease-free equilibrium is not affected by mutations.
 
The main issue is of course whether or not the remaining equilibria $E_j(\mu),\ j=1,...,n$ are non-negative.  
We study this problem next and derive results in terms of the properties of the mutation matrix $Q$. 
 
For the steady-state analysis, we will need the following Lemma which is a  
relevant modification of Theorem A.12 (ii) in \cite{smith-waltman-chem}.  
\begin{lemma}\label{improve} 
Let $M$ be an irreducible square matrix with non-negative off-diagonal entries and 
let $s(M)$ be the stability modulus of $M$. Suppose that there exist $x,r \geq 0$ 
such that $Mx+r=0$. Then the following hold: 
\begin{enumerate} 
\item If $s(M)>0$, then $x=r=0$; 
\item If $s(M)=0$, then $r=0$ and $x$ is a multiple of the positive 
eigenvector of $M$.  
\end{enumerate} 
\end{lemma} 
 
\begin{proof} 
Due to Perron-Frobenius Theorem, $s(M)$ is the principal eigenvalue of $M$. It is 
also the principal eigenvalue of $M'$. Since $M'$ is also irreducible and 
non-negative off-diagonal, there exists $v>0$ such that $M' v =s(M) v$. 
Equivalently, $v'M=s(M)v'$. Hence 
$$ 0 = v'(M x+r)=s(M)v'x+v'r.$$ 
If $s(M)>0$, then both non-negative products $v'x$ and $v'r$ must be zero which implies $x=r=0$. 
If $s(M)=0$, then $v'r=0$ which implies $r=0$.  Hence $Mx=0=s(M)x$ so that $x$ 
is a multiple of the positive eigenvector of $M$. 
\end{proof} 
 
For convenience, we introduce the following notation. We define $A(\mu):=\Gamma^{-1} {\hat N}P(\mu)K$ 
and assume (by renumbering the strains if necessary) that the strains are numbered in such a  
way that the matrix $A(\mu)$ has the lower block-triangular form 
\begin{equation}\label{Amu} 
A(\mu)=\begin{pmatrix} 
A_1(\mu)& 0 & \dots &0 \\ 
\mu B_{2,1} & A_2(\mu) & \dots & 0\\ 
\vdots & \vdots &\ddots & \vdots \\ 
\mu B_{k,1} & \mu B_{k,2}& \dots &\mu A_k(\mu) 
\end{pmatrix}, 
\end{equation} 
where each diagonal block  
$$A_i(\mu)=\diag\biggl(\frac{1}{\bar T_{i1}},\ldots, \frac{1}{\bar T_{is_i}} \biggr)+\mu B_i$$  
is such that $B_i,\ i=1,...,k$ are irreducible with non-negative off-diagonal entries. The off-diagonal 
blocks $B_{i,j},\ i>j$ are non-negative. We note that the diagonal entries of $A(0)$ are a permutation 
of  
$$ 0< \frac{1}{\bar T_{n}} < \frac{1}{\bar T_{n-1}} < \dots < \frac{1}{\bar T_{1}}.$$ 
We say that the strain group $j$ is reachable from strain group $i$ if there exists 
a sequence of indices $i=l_1<l_2<...<l_m=j$ such that all matrices $B_{l_{s+1},l_{s}}$ 
are nonzero. Our first result is as follows: 
\begin{prop}\label{eigenvectors} 
Let the assumptions of Lemma $\ref{invertible}$ hold, 
then the following hold: 
\begin{enumerate} 
\item For all sufficiently small $\mu>0$, matrix $A(\mu)$ admits $n$ distinct 
positive eigenvalues given by 
$$ \frac{1}{\hat T_{n}(\mu)} < \frac{1}{\hat T_{n-1}(\mu)} < \dots < \frac{1}{\hat T_{1}(\mu)},$$ 
such that $\hat T_i(0)=\bar T_i$ for $i=1,...,n$; 
\item Matrix $A(\mu),\ \mu>0$ admits a positive eigenvector $(v_1,v_2,...,v_k)$ if and only 
if $\frac{1}{\hat T_{1}(\mu)}$ is a principal eigenvalue of $A_1(\mu)$, and all strain 
groups $j \geq 2$ are reachable from strain group $1$;  
\item Matrix $A(\mu),\ \mu>0$ admits a non-negative eigenvector $(v_1,v_2,...,v_k)$ for 
each eigenvalue $\frac{1}{\hat T_{r}(\mu)}$ such that 
$\frac{1}{\hat T_{r}(\mu)}$ is a principal eigenvalue of some diagonal block $A_i(\mu)$, 
and $s(A_j(\mu))< \frac{1}{\hat T_{r}(\mu)}$ for all $j=i+1,...,k$ such that strain 
group $j$ is reachable from strain group $i$. The component $v_j$ is 
positive (zero) if group $j$ is reachable (not reachable) from strain group $i$. 
\item All other eigenvectors of $A(\mu),\ \mu>0$ are not sign definite. 
\end{enumerate}  
\end{prop} 
 
\begin{proof} 
The first assertion follows readily because $A(0)$ has $n$ real distinct eigenvalues 
and $A(\mu)$ is continuous (actually, linear) in $\mu$. The continuity of eigenvalues 
with respect to $\mu$ implies that $\hat T_i(0)=\bar T_i$ for $i=1,...,n$. 
 
To prove the second assertion, we begin with sufficiency of the condition. Let $\mu>0$ 
be small and suppose 
that $\frac{1}{\hat T_{1}(\mu)}$ is a principal eigenvalue of $A_1(\mu)$, and all strain 
groups $j \geq 2$ are reachable from strain group $1$. Since $A_1(\mu)$ is 
irreducible with non-negative off-diagonal entries, Perron-Frobenius Theorem implies 
that the eigenvector $v_1$ associated with $\frac{1}{\hat T_{1}(\mu)}$ is positive. 
Since $\frac{1}{\hat T_{1}(\mu)}$ is also the principal eigenvalue of $A(\mu)$, 
it follows that  
$$ s(A_j(\mu)-\frac{1}{\hat T_{1}(\mu)}I)<0, \quad j\geq 2,$$ 
hence $(A_j(\mu)-\frac{1}{\hat T_{1}(\mu)}I)^{-1}<0$ (see e.g. Theorem A.12 (i) in \cite{smith-waltman-chem}). 
The remaining components $v_2,...,v_k$ of the eigenvector satisfy the triangular system 
$$ \begin{array}{ccl} 
0 &= & \mu B_{2,1} v_1 + (A_2(\mu)-\frac{1}{\hat T_{1}(\mu)}I) v_2,\\ 
0 &= & \mu B_{3,1} v_1 + \mu B_{3,2} v_2 + (A_3(\mu)-\frac{1}{\hat T_{1}(\mu)}I) v_3,\\ 
\vdots & \vdots & \vdots\\ 
0 & = & \mu B_{k,1} v_1 + \dots + \mu B_{k,k-1} v_{k-1} + (A_k(\mu)-\frac{1}{\hat T_{1}(\mu)}I) v_k, 
\end{array} 
$$ 
Solving this system recursively, we obtain 
$$ v_j = -(A_j(\mu)-\frac{1}{\hat T_{1}(\mu)}I)^{-1}(\mu B_{j,1} v_1 + \dots + \mu B_{j,j-1} v_{j-1}), 
\quad j=2,...,k.$$ 
Since the strain group $2$ is reachable from strain group $1$, the vector $\mu B_{2,1} v_1 \geq 0$ 
is nonzero. Positivity of the matrix $-(A_2(\mu)-\frac{1}{\hat T_{1}(\mu)}I)^{-1}$ then implies 
that $v_2>0$. By induction on $j$, it follows that $v_j>0$ for all $j=2,...,k$, and hence 
$v=(v_1,v_2,...,v_k)$ is a positive eigenvector.  
 
To prove the converse (the necessary condition),  
let  $v=(v_1,v_2,...,v_k)$ be a positive eigenvector of $A(\mu)$ and let $\frac{1}{\hat T_{q}(\mu)}$ 
be the corresponding eigenvalue. Since $(A_1(\mu)-\frac{1}{\hat T_{q}(\mu)}I)v_1=0$ and $v_1>0$, 
$\frac{1}{\hat T_{q}(\mu)}$ must be the principal eigenvalue of $A_1(\mu)$ (Perron-Frobenius Thm). 
It remains to prove that $s(A_j(\mu)) < \frac{1}{\hat T_{q}(\mu)}$ for all $j\geq 2$. 
Consider $j=2$, and for the sake of contradiction suppose that   
$s(A_j(\mu)-\frac{1}{\hat T_{q}(\mu)}I) \geq 0$. Since the eigenvalues are real and distinct 
for small $\mu>0$, this actually implies $s(A_j(\mu)-\frac{1}{\hat T_{q}(\mu)}I) > 0$. Then we have 
that 
$$ (A_j(\mu)-\frac{1}{\hat T_{q}(\mu)}I) v_2 +\mu B_{2,1} v_1=0$$ 
holds with non-negative vectors $v_2$ and $\mu B_{2,1} v_1$ which are both nonzero. 
By Lemma~\ref{improve}, we have $v_2=0$, a contradiction.  
Hence $s(A_2(\mu)-\frac{1}{\hat T_{q}(\mu)}I) < 0$. Proceeding by induction on $j$, 
we find that $s(A_j(\mu)-\frac{1}{\hat T_{q}(\mu)}I) < 0$ for all $j \geq 2$. 
Therefore, $\frac{1}{\hat T_{q}(\mu)}$ must be the principal eigenvalue of $A(\mu)$, that is, 
$\frac{1}{\hat T_{q}(\mu)}=\frac{1}{\hat T_{1}(\mu)}$. This proves the second assertion. 
 
To prove the third assertion, we again start with sufficient condition. Suppose 
that $\frac{1}{\hat T_{r}(\mu)}$ is a principal eigenvalue of some diagonal block $A_i(\mu)$, 
and $s(A_j(\mu))< \frac{1}{\hat T_{r}(\mu)}$ for all $j=i+1,...,k$ such that strain 
group $j$ is reachable from strain group $i$. It follows immediately that 
all matrices $A_l(\mu)-\frac{1}{\hat T_{r}(\mu)}I,\ l<i$ are nonsingular, and thus 
$v_l=0,\ l<i$. The component $v_i$ is the eigenvector of $A_i(\mu)-\frac{1}{\hat T_{r}(\mu)}I$ 
and it is positive by Perron-Frobenius Theorem. Let $j=i+1$, then one of the following holds. 
If $i+1$ is not reachable from $i$, that is, $B_{i+1,i}=0$ so that 
$$ (A_{i+1}(\mu)-\frac{1}{\hat T_{r}(\mu)}I)v_{i+1}=0$$ 
implies $v_{i+1}=0$ because $A_{i+1}(\mu)-\frac{1}{\hat T_{r}(\mu)}I$ is nonsingular. 
If $i+1$ is reachable from $i$ and $s(A_{i+1}(\mu)-\frac{1}{\hat T_{r}(\mu)}I) < 0$, 
then 
$$ v_{i+1}=-(A_{i+1}(\mu)-\frac{1}{\hat T_{r}(\mu)}I)^{-1} \mu B_{i+1,i} v_i > 0.$$ 
By induction on $j$, it follows that $v_j=0$ for all $j>i$ that are not reachable from $i$ 
and $v_j> 0$ for all $j>i$ that are reachable from $i$. Hence $v=(0,...,0,v_i,v_{i+1},...,v_k)$ 
is a non-negative eigenvector. 
 
Now we prove the necessary condition of the third assertion. Let $v=(v_1,v_2,...,v_k)$ be a 
non-negative eigenvector of $A(\mu)$ associated with eigenvalue $\frac{1}{\hat T_{r}(\mu)}$. 
Let $v_i \geq 0$ be the first nonzero component of $v$, that is, $v=(0,...,0,v_i,...,v_k)$. 
Then $v_i$ satisfies $(A_i(\mu)-\frac{1}{\hat T_{r}(\mu)}I)v_i=0$ hence  
$\frac{1}{\hat T_{r}(\mu)}$ must be an eigenvalue of $A_i(\mu)-\frac{1}{\hat T_{r}(\mu)}I$. 
Moreover, by Perron-Frobenius Theorem, it must be the principal eigenvalue and $v_i>0$. Now consider 
$j=i+1$ and the equation 
$$(A_{i+1}(\mu)-\frac{1}{\hat T_{r}(\mu)}I)v_{i+1} +  \mu B_{i+1,i} v_i=0.$$ 
The vectors $v_{i+1}$ and $\mu B_{i+1,i} v_i$ are non-negative. If 
$s(A_{i+1}(\mu)-\frac{1}{\hat T_{r}(\mu)}I)>0$ then by Lemma~\ref{improve}, 
$\mu B_{i+1,i} v_i=0$. Since $\mu>0$ and $v_i>0$, this implies $B_{i+1,i}=0$. 
Equivalently, $j=i+1$ is not reachable from $i$. An induction argument concludes 
the proof of the third assertion. 
 
The final assertion of this Theorem is a simple one. Let  
$\frac{1}{\hat T_{r}(\mu)}$ be an eigenvalue of $A_i(\mu)$ but not 
the principal eigenvalue and let $v=(v_1,v_2,...,v_n)$ be the 
corresponding eigenvector. Since all eigenvalues of $A(\mu)$ are 
distinct, the matrices $A_{l}(\mu)-\frac{1}{\hat T_{r}(\mu)}I, l<i$ 
are nonsingular so that $v_l=0,\ l<i$. Then $v_i$ must be an 
eigenvector of $A_i(\mu)$ and it cannot be sign definite due to 
Perron-Frobenius theorem. It follows that $v$ is not sign definite. 
\end{proof} 
 
Our second result concerns the existence and the number of non-negative equilibria 
for the systems $(\ref{multic1})-(\ref{multic3})$ and $(\ref{multic4})-(\ref{multic6})$ 
with small $\mu>0$. 
 
\begin{prop}\label{reduce} 
Let the assumptions of Lemma $\ref{invertible}$ hold and suppose that 
the strains are renumbered so that $A(\mu)$ has the form (\ref{Amu}).  
Let $E_j(\mu)=(\hat T_j(\mu),\hat T^*_j(\mu),\hat V_j(\mu))$ denote the nontrivial equilibria of both  
$(\ref{multic1})-(\ref{multic3})$ and $(\ref{multic4})-(\ref{multic6})$ 
for small $\mu>0$. Then  
\begin{enumerate} 
\item $E_j(\mu)$ is positive if and only if $\frac{1}{\hat T_j(\mu)}$ is 
an eigenvalue of $A(\mu)$ with a positive eigenvector $V_j$. 
\item $E_j(\mu)$ is non-negative if and only if $\frac{1}{\hat T_j(\mu)}$ is 
an eigenvalue of $A(\mu)$ with a non-negative eigenvector $V_j$. 
\item $E_j(\mu) \notin \reals^{2n+1}_+$ if and only if  $\frac{1}{\hat T_j(\mu)}$ is an eigenvalue of  
$A(\mu)$ with eigenvector $V_j$ which is not sign-definite. 
 
\end{enumerate} 
\end{prop} 
 
\begin{proof} 
We will prove the Proposition only for system $(\ref{multic1})-(\ref{multic3})$ (the proof for  
$(\ref{multic4})-(\ref{multic6})$ is similar). Observe that the 
equilibrium relation following from $(\ref{multic3})$, can be expressed as 
$\hat T^*_j(\mu)=({\hat N}B)^{-1}\Gamma {\hat V_j}(\mu)$. 
Hence, the signs of the corresponding components of $\hat T^*_j(\mu)$ and ${\hat V_j}(\mu)$ 
are the same. 
Substituting the above expression into $(\ref{multic2})$ and $(\ref{multic3})$, 
we find that ${\hat V_j}(\mu)$ must satisfy 
$$ 
[\Gamma^{-1} {\hat N}P(\mu)K-\frac{1}{{\hat T_j}(\mu)}I]{\hat V_j}(\mu)= 
[A(\mu)-\frac{1}{{\hat T_j}(\mu)}I]{\hat V_j}(\mu)=0. 
$$ 
Thus for each nontrivial equilibrium $E_j(\mu)$, the quantity $\frac{1}{{\hat T_j}(\mu)}$ 
must be an eigenvalue of $A(\mu)$ and ${\hat V_j}(\mu)$ must be a multiple of the 
corresponding eigenvector $V_j$. If $V_j$ is not sign definite, it follows that  
$E_j(\mu) \notin \reals^{2n+1}_+$. 
For all $V_j \geq 0$, the components of $E_j(\mu)$ are uniquely determined via 
$$ \hat V_j(\mu)=\frac{f(\hat T_j(\mu))}{\hat T_j(\mu) k' V_j} V_j, \quad 
\hat T^*_j(\mu)=({\hat N}B)^{-1}\Gamma {\hat V_j}(\mu).$$ 
Hence $E_j(\mu)$ is positive (non-negative) if and only if $V_j$ is positive (non-negative). 
\end{proof} 
 
An immediate corollary to Propositions \ref{eigenvectors} 
and \ref{reduce} is that if the mutation matrix $Q$ is 
irreducible, then $A(\mu)$ is irreducible and  
systems $(\ref{multic1})-(\ref{multic3})$ and $(\ref{multic4})-(\ref{multic6})$ 
with small $\mu>0$ admit a unique positive equilibrium $E_1(\mu)$ and no  
other nontrivial non-negative equilibria. If the 
mutation matrix $Q$ is reducible, then positive equilibrium exists 
if and only the fittest strain (with lowest value $\bar T_1$) belongs 
to strain group 1 and all other strain groups are reachable from group 1, 
meaning that the fittest strain can eventually mutate into any other 
strain. In addition, nontrivial non-negative equilibria which are not positive are  
feasible for $\mu>0$ only if $Q$ is reducible.  
Specifically, if the strains can be numbered according 
to (\ref{Amu}), then at most $k$ nontrivial non-negative equilibria exist. 
One extreme case is when the fittest strain belongs to group $k$, in which 
no positive and only one non-negative equilibrium exist.  
The opposite extreme case is $k=n$ where $A(\mu)$ 
is lower-triangular, the diagonal entries of $A(\mu)$ are arranged in  
decreasing order, and for any pair $i<j$,  strain $j$ is reachable from strain $i$.  
In this case, there is a single positive equilibrium and $n-1$ non-negative equilibria.

\subsection{On uniform strong repellers} 
 
Inspired by Thieme \cite{thieme93}, we make the following definition. 
 
\begin{defi}\label{repellor} 
Consider a system  
\begin{equation} 
\dot x=F(x) \label{sys} 
\end{equation} 
on a compact forward invariant set $K \subset \reals^m$ with a continuous flow $\phi(t,x)$. 
Let $K_0 \subset K$ be a closed forward invariant subset of $K$. Let $d(x,A)$ denote 
the distance from a point $x$ to the set $A$. We say that $K_0$ is a uniform strong repeller in $K$ if 
there exists a $\delta>0$ such that for all solutions $\phi(t,x) \in K \backslash K_0$, 
$\liminf_{t\to\infty} d(\phi(t,x),K_0) \geq \delta$. 
\end{defi} 
 
\begin{stel}\label{patrickdoeshofbauer} 
Let $\Pi: K \to \reals^+$ be a continuously differentiable function such that  
$\Pi(x)=0$ if and only if $x\in K_0$.  
Suppose there exists a lower semi-continuous  
function $\psi: K \to \reals$ such that 
\begin{equation} 
\frac{\dot \Pi}{\Pi}=\psi,\quad \forall x\in K\backslash K_0. 
\label{logdot}\end{equation} 
Suppose that the following condition holds  
$$ (H) \quad \forall x \in K_0, \  \exists T>0:  \ \langle \psi(\phi(T,x)) \rangle >0.$$ 
Then $K_0$ is a uniform strong repellor in $K$. 
\end{stel} 
 
\begin{proof}  
{\it Step 1}.  
Note that by lower semi-continuity of $\psi$ and continuity of $\phi$, for every $p\in K_0$  
we can  find an open set $U_p$ containing $p$, and a  
lower semi-continuous map $T_p:U_p \rightarrow (0,+\infty)$ so that for every $q \in U_p$, (H) holds with $x=q$ and $T=T_p(q)$.  
Choose for every $p\in K_0$ a non-empty open set $V_p$ with ${\bar V_p}\subset U_p$.  
Then by lower semi-continuity of each map  
$T_p$ and compactness of $\bar V_p$,  
$$  
\inf_{q\in {\bar V_p}}T_p(q) >0  
$$  
is achieved in ${\bar V_p}$.  
Since $\cup_{p\in K_0}V_p$ is an open cover of $K_0$, we may choose a finite open subcover  
$\cup_{i=1,\dots ,n}V_{p_i}$. Let $\tau_i=\inf_{q\in {\bar V_{p_i}}}T_{p_i}(q)>0$ and set  
$$  
\tau=\min_{i=1,\dots,n}\tau_i>0.  
$$  
Note that for every $p\in K_0$, there is a $T\geq \tau$ so that (H) holds with $x=p$. That is, $\tau$ is a  
uniform (in $K_0$) lower bound for $T$'s for which (H) holds.  
 
{\it Step 2}. Let $h>0$ be given. Define  
\begin{equation}\label{layer}  
U_h=\{x\in K\;|\;\exists \,T> \tau:\langle \psi(\phi(T,x))\rangle > h\}  
\end{equation}  
We claim that $U_h$ is open.  
 
Fix $z\in U_h$. Then there is some $T> \tau$ so that  
$$  
\epsilon:=\langle \psi(\phi(T,z))\rangle -h>0.  
$$  
Then by continuity of $\phi$  
and lower semi-continuity of $\psi$  
(and therefore uniform lower semi-continuity of $\psi$ on compact sets), it follows that there  
exists an open set $W_z$  
containing $z$ such that for all $z'\in W_z$ holds that  
\begin{equation}\label{bounds}  
\psi(\phi(t,z')>\psi(\phi(t,z))-\epsilon,\;\; \forall t\in [0,T].  
\end{equation}  
Now since  
$$  
\langle \psi(\phi(T,z))\rangle=\epsilon +h,  
$$  
it follows from $(\ref{bounds})$ that for all $z'\in W_z$:  
$$  
\langle \psi(\phi(T,z'))\rangle>h,  
$$  
and thus that $W_z\subset U_h$, establishing our claim.  
 
{\it Step 3}. Define $T_h:U_h\rightarrow [\tau,+\infty)$ as  
$$  
T_h(z)=\inf\{T>\tau\;|\; \langle \psi(\phi(T,z)) \rangle>h\}.  
$$  
We claim that $T_h$ is upper semi-continuous.  
 
Fix $z\in U_h$ and let $\epsilon'>0$ be given. Then there is some $T>\tau$ so that  
$$  
\langle \psi(\phi(T,z))\rangle>h,  
$$  
so that  
\begin{equation}\label{first}  
T<T_h(z)+\epsilon'  
\end{equation}  
By the argument in Step 2, there is some open set $W_z$ containing $z$, such that for all  
$z'\in W_z$ holds that:  
$$  
\langle \psi(\phi(T,z'))\rangle>h,  
$$  
and thus that for all $z'\in W_z$:  
\begin{equation}\label{second}  
T_h(z')\leq T  
\end{equation}  
Our claim follows by combining $(\ref{first})$ and $(\ref{second})$.  
 
{\it Step 4}. The nested family $\{U_h\}_{h>0}$ is decreasing (under set inclusion),  
and forms an open cover of $K_0$. Hence,  
there is some ${\bar h}$ so that $U_{{\bar h}}$ covers $K_0$.  
Since ${\tilde K}:=K \setminus U_{{\bar h}}$ is  
compact, and $\Pi$ is continuous, $\Pi$ attains its minimal value $m>0$ on ${\tilde K}$.  
Choose $p\in (0,m)$ and define:  
$$  
I_p=\{z\in K\;|\;\Pi(z)\in (0,p]\}.  
$$  
Then $I_p \subset U_{{\bar h}}$.  
 
{\it Step 5}. We claim that every forward solution starting in $I_p$, eventually leaves $I_p$, that is:  
$$  
\forall z\in I_p, \exists t_z>0:\;\; \phi(t_z,z)\notin I_p.  
$$  
By contradiction, if $\phi(t,z)\in I_p$ for all $t\geq 0$, then $\phi(t,z)\in U_{{\bar h}}$ for all $t\geq 0$, and thus:  
$$  
\exists T_t\geq \tau:\;\; \frac{1}{T_t}\int_t^{t+T_t}\psi(\phi(s,z))ds > {\bar h}.  
$$  
Then integrating equation (\ref{logdot}) from $t$ to $t+T$ yields that:  
$$  
\ln\left(\frac{\Pi(\phi(t+T_t,z))}{\Pi(\phi(t,z))} \right) >{\bar h}T_t,  
$$  
and thus that  
\begin{equation}\label{div-seq}  
\Pi(\phi(t+T_t,z))>e^{{\bar h}T_t}\Pi(\phi(t,z)).  
\end{equation}  
Set $t_0=0$ and $t_k=t_{k-1}+T_{t_{k-1}}$ for $k=1,2,\dots$. Since each $T_{t_k}\geq \tau >0$ it follows that  
$t_k\rightarrow \infty$. Then by $(\ref{div-seq})$ and since $t_k\geq \tau$ for all $k$, we have that:  
$$  
\Pi(\phi(t_k,z))>e^{{\bar h}T_{t_{k-1}}}\Pi(\phi(t_{k-1},z))>e^{k \tau}\Pi(z),  
$$  
so that $\Pi(\phi(t_k,z))\rightarrow \infty$ as $k\rightarrow \infty$.  
This contradicts boundedness of $\Pi$ on the compact set $K$.  
 
{\it Step 6}. Let  
$$  
{\tilde I}_p=I_p \cup K_0.  
$$  
We will show that there is some $q\in (0,p)$ so that forward solutions starting outside ${\tilde I}_p$, never  
reach $I_q$, that is:  
$$  
\exists q\in (0,p):\;\; z\notin {\tilde I}_p\Rightarrow \phi(t,z)\notin I_q,\;\; \forall t\geq 0.  
$$  
Consider a forward solution $\phi(t,z)$ with $z\notin {\tilde I}_p$. If  
$\phi(t,z)\notin {\tilde I}_p$ for all $t\geq 0$,  
then we are done since ${\tilde I}_q \subset {\tilde I}_p$, so let us 
assume that for some $t_z>0$, holds that $\phi(t_z,z)\in {\tilde I}_p$.  
Denote the first time this  
happens by $t_0$:  
$$  
t_0=\min \{t>0\;|\; \phi(t,z)\in {\tilde I}_p\}.  
$$  
Set $z^*=\phi(t_0,z)$ and note that $\Pi(z^*)=p$. Denote $\inf_{z\in K_0}\psi(z)$ by $m'$.  
If $m'\geq 0$, then (\ref{logdot}) implies that $\Pi(\phi(t,z^*))\geq \Pi(z^*)=p$ for all $t\geq 0$, so  
that we're done. If on the the other hand $m'<0$, we first define  
$$  
{\bar T}=\max_{z\in {\tilde I}_p} T_h(z) (\geq \tau>0).  
$$  
Notice that this maximum is indeed achieved on the compact set ${\tilde I}_p$, since $T_h$  
is upper semi-continuous.  
Now we define  
$$  
q=pe^{m'{\bar T}},  
$$  
and notice that $q$ is independent of the chosen solution $z(t)$. We  
will show that for this choice of $q$, our claim is established.  
 
We have that:  
$$  
\forall t\in (0,{\bar T}):\;\; \frac{1}{t}\int_0^t\psi(\phi(s,z^*))ds\geq m',  
$$  
and thus by (\ref{logdot}) that  
\begin{equation}\label{one-hand}  
\forall t\in (0,{\bar T}):\;\; \Pi(\phi(t,z^*))\geq \Pi(z^*)e^{m't}>q,  
\end{equation}  
which implies that during the time interval $(0,{\bar T})$, the solution $\phi(t,z^*)$ has not reached $I_q$.  
On the other hand, during that same time interval $(0,{\bar T})$, the solution $\phi(t,z^*)$ must  
have left ${\tilde I}_p$.    
If this were not the case, then by the argument in Step 5, there would be some  
$T^*\in [\tau , {\bar T})$ so that  
$$  
\Pi(\phi(T^*,z^*))\geq \Pi(z^*)e^{{\bar h}T}>p,  
$$  
and thus that $\phi(T^*,z^*)\notin {\tilde I}_p$, a contradiction to our assumption.  
This process can be repeated  
iteratively  
and leads to the conclusion that the forward solution $\phi(t,z)$ which did not start in ${\tilde I}_p$, will  
never reach $I_q$.  
 
So far we 
have shown that for any solution $\phi(t,x) \notin K_0$, inequality $\Pi(\phi(t,x)) \geq q>0$ 
for all sufficiently large $t$. The sets $K_0=\Pi^{-1}(\{0\})$ and $\Pi^{-1}([q,+\infty)) \cap K$ 
are compact and disjoint. Therefore, there exists $\delta>0$ such that $d(\phi(t,x),K_0) \geq \delta$ 
for all $x\notin K_0$ and all sufficiently large $t$. 
\end{proof}

\subsection{Global stability for small $\mu>0$} 
The following Lemmas will  
be used to prove global stability of the positive 
 equilibrium for small $\mu>0$. 
 
\begin{lemma}\label{liminf} 
Let $a: \reals^m \to \reals^n$ be continuous and let $b \in {\rm int}(\reals^n_+)$. Let 
$f: \reals^m \times (\reals^n_{+} \backslash \{0\}) \to \reals$ be defined as 
$$ f(x,y) =\frac{a'(x)y}{b'y}.$$ 
Then   
\begin{equation} \liminf_{x\to x_0, y\to 0+} f(x,y)=\min_{i\in\{1,...,n\}} \frac{a_i(x_0)}{b_i}, 
\label{LIM}\end{equation} 
furthermore,  
if we define $f(x,0)=\min_{i\in\{1,...,n\}} \frac{a_i(x)}{b_i},$ 
then $f(x,y)$ becomes a lower semi-continuous function on $\reals^m \times \reals^n_{+}$ 
whose restriction on $\reals^m \times \{0\}$ is continuous.   
\end{lemma} 
 
\begin{proof} 
Extending the function $f(x,y)$ by defining $f(x_0,0)=\liminf_{x\to x_0, y\to 0+} f(x,y)$ 
clearly produces a lower semi-continuous function. Furthermore, since $a(x)$ 
is continuous, the function $\min_{i\in\{1,...,n\}} \frac{a_i(x)}{b_i}$ is 
continuous as well. So it remains to show that (\ref{LIM}) holds.  
 
Without loss of generality, we may assume that $\min_{i\in\{1,...,n\}} \frac{a_i(x_0)}{b_i} 
=\frac{a_1(x_0)}{b_1}$.  Setting $x=x_0$ and $y_2=y_3=...=y_n=0$ and letting $y_1\to 0^+$, 
we find that $f(x_0,y_1,0,...,0) \to \frac{a_1(x_0)}{b_1}.$ Hence,  
$\liminf_{x\to x_0, y\to 0+} f(x,y)\leq \frac{a_1(x_0)}{b_1}.$ 
We also observe that as long as $y\in \reals^n_+ \backslash \{0\}$, the value 
$$ \frac{a'(x)y}{b'y} = \sum_{i=1}^n \frac{a_i(x)}{b_i} \frac{b_i y_i}{b_1 y_1+ \cdots + b_n y_n}$$ 
is a convex linear combination of the values $\frac{a_i(x)}{b_i},\ i=1,...,n$. 
By continuity of $a(x)$, for any $\varepsilon >0$ there exists $\delta>0$ 
such that $\forall i\in\{1,...,n\}$ and $\forall x \in B_{\delta}(x_0)$, 
we have $a_i(x) > a_i(x_0)- \varepsilon b_i$. Hence, for all $x \in B_{\delta}(x_0)$ 
and for all  $y\in \reals^n_+\backslash \{0\}$, $f(x,y) \geq \frac{a_1(x_0)}{b_1} -\varepsilon$. 
We have established that  
$$\frac{a_1(x_0)}{b_1} \geq \liminf_{x\to x_0, y\to 0+} f(x,y)\geq \frac{a_1(x_0)}{b_1}-\varepsilon.$$ 
Since $\varepsilon>0$ is arbitrary, (\ref{LIM}) follows. 
\end{proof} 
 
\begin{lemma}\label{lowerbound}
Suppose that (\ref{T's}) holds. Then 
there exist $\eta,\mu_0>0$ such that  
$$\liminf_{t\to\infty} 1' V(t) \geq \eta >0$$ 
for any $\mu\in[0,\mu_0]$ and for any solution of $(\ref{multic1})-(\ref{multic3})$ and  
$(\ref{multic4})-(\ref{multic6})$ with $1' V(t)>0$. 
\end{lemma} 
 
\begin{proof} 
We will prove the claim for system $(\ref{multic1})-(\ref{multic3})$ (the proof for 
$(\ref{multic4})-(\ref{multic6})$ is similar). The proof consists of two parts.  
We first show that there exist  $\eta_0,\mu_0>0$ such 
 that $\liminf_{t\to\infty} 1'(T^*(t)+V(t)) \geq \eta_0 >0$ for all solutions with $T^*(t),V(t) \not=0$. 
 
We choose $n$ positive numbers $\tilde N_i$ so that $\frac{\gamma_i}{k_i T_0} < \tilde N_i<N_i$ 
for all $1 \leq i \leq n$. This is possible since we assume  
$\bar T_i =\frac{\gamma_i}{k_i N_i} < T_0$. Let $v=(\tilde N,1)$. It follows that 
$$ v'\begin{pmatrix} -B & K T_0\cr \hat N B & -\Gamma\cr\end{pmatrix}= 
\biggl(b_1(N_1-\tilde N_1),...,b_n(N_n-\tilde N_n), k_1 T_0 \tilde N_1 - \gamma_1,..., 
k_n T_0 \tilde N_n - \gamma_n\biggr)$$ 
is a positive vector. By continuity, there exists a $\mu_0>0$ such that 
$$v'M(T,\mu),\quad {\rm where}\  
M(T,\mu):=\begin{pmatrix} -B & P(\mu) K T\cr \hat N B & -\Gamma\cr\end{pmatrix}$$ 
is a positive vector for all $\mu \in [0,\mu_0]$.  
 
Consider a system  
\begin{eqnarray} 
{\dot T}&=&f(T)-k'VT,\ T \in \reals_+\label{pers1}\\ 
{\dot T^*}&=&P(\mu)KVT-B T^*,\ T^* \in \reals^n_+,\label{pers2}\\ 
{\dot V}&=&{\hat N}B T^*-\Gamma V,\ V\in \reals^n_+,\label{pers3}\\ 
{\dot \mu}&=& 0,\ \mu \in [0,\mu_0].\label{pers4} 
\end{eqnarray} 
Let $K'$ be the forward invariant compact set for $(\ref{multic1})-(\ref{multic3})$  
established in Lemma~\ref{bounded} and define $K=K'\times[0,\mu_0]$. It is clear that 
$K$ is compact and forward invariant under $(\ref{pers1})-(\ref{pers4})$ 
The set 
$$K_0:=\left([0,T_0]\times \{0\} \times \{0\} \times [0,\mu_0]\right) \cap K$$ is 
clearly a compact forward invariant subset of $K$.  
 
Let $\Pi(T^*,V):=v'(T^*,V)$. The function $\Pi$ is clearly 
smooth, zero on $K_0$, and positive on $K\backslash K_0$. Furthermore, 
$$ \frac{\dot \Pi}{\Pi}= \psi := 
\frac{v'M(T,\mu)(T^*,V)}{v'(T^*,V)}$$ 
is lower semi-continuous on $K$ by Lemma~\ref{liminf} once we define the value of $\psi$ on $K_0$ 
as 
$$ \psi(T,\mu)=\min_{i=1,...,n} 
\frac{v'M(T,\mu)_i}{v_i} .$$ 
We note that the function $\psi(T,\mu)$ is continuous in $(T,\mu)$. 
Since all solutions of $(\ref{pers1})-(\ref{pers4})$ in $K_0$ have the property that  
$\lim_{t\to\infty}T(t)=T_0$, it implies that $\psi(T(t),\mu)>0$ for all sufficiently large $t$. 
Hence by Theorem~\ref{patrickdoeshofbauer}, the set $K_0$ is a uniform strong repeller in $K$. 
If we use the $L^1$-norm of $(T^*,V)$ as the distance function to $K_0$, we find that there 
exists an $\eta_0>0$ such that 
$$ \liminf_{t\to\infty} 1'(T^*+V) \geq \eta_0$$ 
for all solutions of $(\ref{pers1})-(\ref{pers4})$ in $K\backslash K_0$.  
 
To complete the proof, we need to show that there exists $\eta>0$ such that 
$\liminf_{t\to\infty} 1'V(t) \geq \eta >0$ for all solutions with $1'V(t)>0$.  
Observe that $1'V(t)>0$ implies that $1'T^*(t)>0$. Hence by the result of 
part one, we have that $\liminf_{t\to\infty} 1'(T^*(t)+V(t)) \geq \eta_0 >0$, or 
equivalently, $1'T^*(t) > \eta_0/2 - 1'V(t)$ for all sufficiently large $t$. We substitute this 
inequality into (\ref{multic3}) and find that 
$$ 1'\dot V \geq A_0 \biggl(\frac{\eta_0}{2} - 1'V(t)\biggr)-A_1 1'V(t), \quad A_0:=\min_{i} (N_i \beta_i)>0, 
\ A_1:=\max_{i} (\gamma_i) >0$$ 
holds for large $t$. It follows immediately that 
$$ \liminf_{t\to\infty} 1'V(t) \geq \eta=\frac{\eta_0 A_0}{2(A_0+A_1)}>0.$$ 
\end{proof} 

\begin{lemma}\label{minilemma}
Let
$$ \sigma(x,y,z):= x+ y + \frac{z}{xy}-3 z^\frac{1}{3}.$$
Then for any $z_0,M>0$, there exists $\delta>0$ such that
$\sigma(x,y,z)>M$ for all $0<x<\delta$, all $y>0$, and all $z>z_0$.
\end{lemma}

\begin{proof}
Observe that the minimum of the function $\sigma(x,\cdot,z)$ on the set $y\in (0,+\infty)$ is
achieved at $y=\sqrt{z/x}$. Hence for all $y>0$, it holds that 
$$f(x,y,z) \geq f(x,\sqrt{\frac{z}{x}},z)=x+ 2 \sqrt{\frac{z}{x}} -3 z^\frac{1}{3}.$$
Let $z_0>0$ and define
$$\delta:=\frac{4 z_0}{\left( M+ 3 z_0^\frac{1}{3} \right)^2}.$$
Then for all $0<x<\delta$, all $y>0$, and all $z>z_0$, it holds that
$$f(x,y,z) \geq 2 \sqrt{\frac{z}{x}} -3 z^\frac{1}{3}
=z^\frac{1}{2} \left( \frac{2}{x^{\frac{1}{2}}}-3 z^{-\frac{1}{6}} \right) >
z_0^\frac{1}{2} \left( \frac{2}{\delta^\frac{1}{2}}-3 z_0^{-\frac{1}{6}} \right)=M.$$
\end{proof}
 
\begin{stel}\label{compactset} 
Let $K$ be the absorbing compact set established  
in Lemma~\ref{bounded}, and let  
$$ U=\{ (T,T^*,V) \in \reals^{2n+1}_+ | T,T^*_1,V_1>0 \}.$$ 
 Suppose that ${\bf (C)}$ holds with ${\bar T_1}$ instead of ${\bar T}$. 
Then there exist $\mu_1>0$ and a compact set $K_{\delta} \subset U$ such that for  
any $\mu\in[0,\mu_1]$ and for  any solution of $(\ref{multic1})-(\ref{multic3})$ or  
$(\ref{multic4})-(\ref{multic6})$ in $U$, there exists a $t_0>0$ such that  
$ (T(t),T^*(t),V(t)) \in K_{\delta}$ for  
all $t >t_0$.  
\end{stel} 
 
\begin{proof} Both for system $(\ref{multic1})-(\ref{multic3})$ and $(\ref{multic4})-(\ref{multic6})$,   
the  proof will be based on the same Lyapunov function 
$$ 
W=\int_{{\bar T}^1}^T \left(1-\frac{{\bar T}^1}{\tau}\right) d \tau +\int_{{\bar T^*}_1}^{T^*_1} \left(1-\frac{{\bar T^*}_1}{\tau}\right)d \tau+ 
\frac{1}{N_1}\int_{{\bar V}_1}^{V_1} \left(1-\frac{{\bar V}_1}{\tau}\right) d \tau+\sum_{i>1}T^*_i+\frac{1}{N_i}V_i 
$$ 
that we used to show competitive exclusion with $\mu=0$.  
 
{\bf Case 1}: System $(\ref{multic1})-(\ref{multic3})$. 
 
Computing $\dot W$ for system $(\ref{multic1})-(\ref{multic3})$, we obtain after some simplifications 
\begin{eqnarray*} 
{\dot W}&=&(f(T)-f({\bar T}^1))\left(1-\frac{{\bar T}^1}{T} \right)- 
\beta_1 {\bar T^*}_1\left[\frac{{\bar T}^1}{T}+\frac{{\bar T^*}_1V_1T}{T^*_1{\bar V}_1{\bar T}^1}+ 
\frac{{\bar V}_1T^*_1}{V_1{\bar T^*}_1}-3\right]\\ 
&&-\sum_{i=2}^nk_iV_i({\bar T}^i-{\bar T}^1) +\frac{T^*_1-\bar T^*_1}{T^*_1} \mu  
\sum_{j=1}^n q_{1j} k_j V_j T + \mu \sum_{i=2}^n \sum_{j=1}^n q_{ij} k_j V_j T. 
\end{eqnarray*} 
Recombining the terms, we further obtain 
\begin{eqnarray*} 
{\dot W}&=&(f(T)-f({\bar T}^1))\left(1-\frac{{\bar T}^1}{T} \right)- 
\beta_1 {\bar T^*}_1\left[\frac{{\bar T}^1}{T}+\frac{{\bar T^*}_1V_1T}{T^*_1{\bar V}_1{\bar T}^1}+ 
\frac{{\bar V}_1T^*_1}{V_1{\bar T^*}_1}-3\right]\\ 
&&-\sum_{i=2}^nk_iV_i({\bar T}^i-{\bar T}^1) -\frac{\bar T^*_1}{T^*_1} \mu  
\sum_{j=1}^n q_{1j} k_j V_j T + \mu \sum_{i=1}^n \sum_{j=1}^n q_{ij} k_j V_j T. 
\end{eqnarray*} 
We note that  
$$ \sum_{i=1}^n \sum_{j=1}^n q_{ij} k_j V_j T = 
\sum_{j=1}^n \biggl( \sum_{i=1}^n q_{ij}\biggr) k_j V_j T=0$$ 
since all column sums of $Q$ are zero. Hence, 
\begin{eqnarray*} 
{\dot W}&=&(f(T)-f({\bar T}^1))\left(1-\frac{{\bar T}^1}{T} \right)- 
\beta_1 {\bar T^*}_1\left[\frac{{\bar T}^1}{T}+\frac{{\bar T^*}_1V_1T}{T^*_1{\bar V}_1{\bar T}^1}+ 
\frac{{\bar V}_1T^*_1}{V_1{\bar T^*}_1}-3\right]\\ 
&&-\sum_{i=2}^nk_iV_i({\bar T}^i-{\bar T}^1) - \frac{\bar T^*_1}{T^*_1} \mu q_{11} k_1 V_1 T 
- \frac{\bar T^*_1}{T^*_1}\mu \sum_{j=2}^n q_{1j} k_j V_j T. 
\end{eqnarray*} 
We rewrite $\dot W$ as 
\begin{eqnarray*} 
{\dot W}&=&(f(T)-f({\bar T}^1))\left(1-\frac{{\bar T}^1}{T} \right)- 
\beta_1 {\bar T^*}_1\left[\frac{{\bar T}^1}{T}+(1+q_{11}\mu)\frac{{\bar T^*}_1V_1T}{T^*_1{\bar V}_1{\bar T}^1}+ 
\frac{{\bar V}_1T^*_1}{V_1{\bar T^*}_1}-3(1+q_{11}\mu)^{1/3}\right]\\ 
&& + 3\beta_1 {\bar T^*}_1(1-(1+q_{11}\mu)^{1/3})-\sum_{i=2}^nk_iV_i({\bar T}^i-{\bar T}^1)  
- \frac{\bar T^*_1}{T^*_1}\mu \sum_{j=2}^n q_{1j} k_j V_j T. 
\end{eqnarray*} 
Note that the last term of $\dot W$ is non-positive, hence\footnote{Incidentally, 
if $q_{11}=0$, we obtain global stability of the 
boundary equilibrium $E_1$ for all $\mu>0$.}  
\begin{eqnarray*} 
{\dot W}&\leq &(f(T)-f({\bar T}^1))\left(1-\frac{{\bar T}^1}{T} \right)- 
\beta_1 {\bar T^*}_1\left[\frac{{\bar T}^1}{T}+(1+q_{11}\mu)\frac{{\bar T^*}_1V_1T}{T^*_1{\bar V}_1{\bar T}^1}+ 
\frac{{\bar V}_1T^*_1}{V_1{\bar T^*}_1}-3(1+q_{11}\mu)^{1/3}\right]\\ 
&& + 3\beta_1 {\bar T^*}_1(1-(1+q_{11}\mu)^{1/3})-\sum_{i=2}^nk_iV_i({\bar T}^i-{\bar T}^1).  
\end{eqnarray*} 
By Lemma~\ref{lowerbound}, 
 there exist $\eta,\mu_a>0$ such that $1'V(t) > \eta$ for all $\mu \in [0,\mu_a]$ 
 and all sufficiently large $t$.  
Let $\alpha=\min_{i \geq 2} k_i({\bar T}^i-{\bar T}^1)>0$, then 
$$ \sum_{i=2}^nk_iV_i({\bar T}^i-{\bar T}^1) \geq \alpha \sum_{i=2}^n V_i \geq 
\alpha (\eta- V_1).$$ 
Thus, by shifting time forward if necessary, we have the  
inequality 
 
\begin{eqnarray*} 
{\dot W}&\leq &(f(T)-f({\bar T}^1))\left(1-\frac{{\bar T}^1}{T} \right)- 
\beta_1 {\bar T^*}_1\left[\frac{{\bar T}^1}{T}+(1+q_{11}\mu)\frac{{\bar T^*}_1V_1T}{T^*_1{\bar V}_1{\bar T}^1}+ 
\frac{{\bar V}_1T^*_1}{V_1{\bar T^*}_1}-3(1+q_{11}\mu)^{1/3}\right]\\ 
&& + 3\beta_1 {\bar T^*}_1(1-(1+q_{11}\mu)^{1/3})-\alpha \eta + \alpha V_1.  
\end{eqnarray*} 
Let $\mu_b>0$ be such that for all $\mu \in [0,\mu_b]$ 
, 
$$ 1+q_{11} \mu \in \left[\frac{1}{2},1\right], \quad 3\beta_1 {\bar T^*}_1(1-(1+q_{11}\mu)^{1/3})-\alpha \eta 
\leq -\frac{\alpha \eta}{2}.$$ 
Let $\mu_1=\min(\mu_a,\mu_b)$ and choose sufficiently large  
$L>0$ so that  
$$3\beta_1 {\bar T^*}_1(1-(1+q_{11}\mu)^{1/3})-\alpha \eta + \alpha V_1<L$$ for all solutions 
of $(\ref{multic1})-(\ref{multic3})$ in $K$ and all $\mu \in [0,\mu_1]$.  
For any $\mu \in [0,\mu_1]$, we have that 
\begin{eqnarray*} 
{\dot W}&\leq &(f(T)-f({\bar T}^1))\left(1-\frac{{\bar T}^1}{T} \right)- 
\beta_1 {\bar T^*}_1\left[\frac{{\bar T}^1}{T}+(1+q_{11}\mu)\frac{{\bar T^*}_1V_1T}{T^*_1{\bar V}_1{\bar T}^1}+ 
\frac{{\bar V}_1T^*_1}{V_1{\bar T^*}_1}-3(1+q_{11}\mu)^{1/3}\right]\\ 
&& -\frac{\alpha \eta}{2} + \alpha V_1,  
\end{eqnarray*} 
where the first two terms are non-positive and $1+q_{11}\mu \in \left[\frac{1}{2},1\right]$. 
Inspecting the first term in $\dot W$,  
we find that there exists $\delta_0>0$ such that  
$$ (f(T)-f({\bar T}^1)\left(1-\frac{{\bar T}^1}{T} \right)<-(L+1)$$ 
for all $T<\delta_0$ and all $\mu \in [0,\mu_1]$. Now we inspect the the second 
term in $\dot W$. Using  Lemma~\ref{minilemma}
with 
$$ x=\frac{{\bar V}_1T^*_1}{V_1{\bar T^*}_1}, \quad
y= \frac{{\bar T}^1}{T}, \quad
z=1+q_{11}\mu, \quad z_0=\frac{1}{2},$$ we conclude that
 there exists $\delta_1>0$ such that  
$$ -\beta_1 {\bar T^*}_1\left[\frac{{\bar T}^1}{T}+(1+q_{11}\mu)\frac{{\bar T^*}_1V_1T}{T^*_1{\bar V}_1{\bar T}^1}+ 
\frac{{\bar V}_1T^*_1}{V_1{\bar T^*}_1}-3(1+q_{11}\mu)^{1/3}\right]<-(L+1)$$ 
for all $\frac{T^*_1}{V_1}<\delta_1$ and all $\mu \in [0,\mu_1]$. Finally, 
there exists $\delta_2>0$ such that  
$-\frac{\alpha\eta}{2}+ \alpha V_1 <-\frac{\alpha \eta}{4}$ 
for all $V_1<\delta_2$ and all $\mu \in [0,\mu_1]$. 
Let  
$$\hat K_{\delta}=\{ (T,T^*,V) \in K \cap U | T \geq \delta_0, V_1 \geq \delta_2, 
T^*_1 \geq \delta_1 V_1\}.$$ 
Consider $(T,T^*,V) \in (K \cap U) \backslash \hat K_{\delta}$ 
and let $\mu \in [0,\mu_1]$, then at least one of the following holds: 
\begin{itemize} 
\item $T <\delta_0$, in which case $ \dot W \leq -(L+1) + L \leq -1$; 
\item $T^*_1/V_1 < \delta_1$, in which case $ \dot W \leq -(L+1) + L \leq -1$; 
\item $ V_1 < \delta_2$, in which case $\dot W \leq  -\frac{\alpha \eta}{4}$; 
\end{itemize} 
Hence, for all $(T,T^*,V) \in (K \cap U) \backslash \hat K_{\delta}$ 
and all $\mu \in [0,\mu_1]$, we have 
$$ \dot W \leq -\min(1,\frac{\alpha \eta}{4})<0.$$ We postpone the rest of the proof until we have showed  
that a similar inequality holds for system $(\ref{multic4})-(\ref{multic6})$. 
 
{\bf Case 2}: System $(\ref{multic4})-(\ref{multic6})$. 
 
Computing $\dot W$ for system $(\ref{multic4})-(\ref{multic6})$, we obtain after some simplifications 
\begin{eqnarray*} 
{\dot W}&=&(f(T)-f({\bar T}^1))\left(1-\frac{{\bar T}^1}{T} \right)- 
\beta_1 {\bar T^*}_1\left[\frac{{\bar T}^1}{T}+\frac{{\bar T^*}_1V_1T}{T^*_1{\bar V}_1{\bar T}^1}+ 
\frac{{\bar V}_1T^*_1}{V_1{\bar T^*}_1}-3\right]\\ 
&&-\sum_{i=2}^nk_iV_i({\bar T}^i-{\bar T}^1) + 
\mu\left(\frac{V_1-{\bar V_1}}{V_1} \right)\sum_{j=1}^n q_{1j}\frac{N_j}{N_1}\beta_j T^*_j+ 
\mu\sum_{i=2}^n\sum_{j=1}^n q_{ij}\frac{N_j}{N_i}\beta_j T^*_j. 
\end{eqnarray*} 
Note that the $\mu$ dependent terms can be rearranged as follows: 
$$ 
\mu\left(\sum_{i=1}^nq_{ii} \beta_i T^*_i - \frac{{\bar V_1}}{V_1}\sum_{j=2}^nq_{1j}\frac{N_j}{N_1}\beta_j T^*_j\right) 
+\mu\left(\sum_{j=2}^nq_{1j}\frac{N_j}{N_1}\beta_j T^*_j +\sum_{i=2}^n \sum_{j\neq i}^n q_{ij}\frac{N_j}{N_i}\beta_j T^*_j\right) 
-\mu\frac{{\bar V_1}}{V_1}q_{11}\beta_1 T^*_1. 
$$ 
In the above the first term is non-positive, and the second term can be re-written as follows: 
$$ 
\mu\sum_{i=1}^n\alpha_iT^*_i, 
$$ 
for suitable $\alpha_i\geq 0$, and the third term will be absorbed in the square bracket $[ \;\;]$ term in ${\dot W}$.  
We find that 
\begin{eqnarray*} 
{\dot W}&\leq&(f(T)-f({\bar T}^1))\left(1-\frac{{\bar T}^1}{T} \right)- 
\beta_1 {\bar T^*}_1\left[\frac{{\bar T}^1}{T}+\frac{{\bar T^*}_1V_1T}{T^*_1{\bar V}_1{\bar T}^1}+ 
(1+q_{11}\mu)\frac{{\bar V}_1T^*_1}{V_1{\bar T^*}_1}-3(1+q_{11}\mu)^{1/3}\right]\\ 
&& + 3\beta_1 {\bar T^*}_1(1-(1+q_{11}\mu)^{1/3})+\mu \sum_{i=1}^n \alpha_i T^*_i -\sum_{i=2}^nk_iV_i({\bar T}^i-{\bar T}^1).  
\end{eqnarray*} 
By Lemma~\ref{lowerbound}, there exist $\eta,\mu_a>0$ such that $1'V(t) > \eta$ for all $\mu \in [0,\mu_a]$  
and all sufficiently large $t$.  
Let $\alpha=\min_{i \geq 2} k_i({\bar T}^i-{\bar T}^1)>0$, then 
$$ \sum_{i=2}^nk_iV_i({\bar T}^i-{\bar T}^1) \geq \alpha \sum_{i=2}^n V_i \geq 
\alpha (\eta- V_1).$$ 
Thus, by shifting time forward if necessary, we have the inequality 
\begin{eqnarray*} 
{\dot W}&\leq&(f(T)-f({\bar T}^1))\left(1-\frac{{\bar T}^1}{T} \right)- 
\beta_1 {\bar T^*}_1\left[\frac{{\bar T}^1}{T}+\frac{{\bar T^*}_1V_1T}{T^*_1{\bar V}_1{\bar T}^1}+ 
(1+q_{11}\mu)\frac{{\bar V}_1T^*_1}{V_1{\bar T^*}_1}-3(1+q_{11}\mu)^{1/3}\right]\\ 
&& + 3\beta_1 {\bar T^*}_1(1-(1+q_{11}\mu)^{1/3})+\mu \sum_{i=1}^n \alpha_i T^*_i -\alpha \eta + \alpha V_1. 
\end{eqnarray*} 
Since solutions are in the compact set $K$ for sufficiently large times, there is some $\mu'_a>0$ such that  
$$ 
\mu\sum_{i=1}^n \alpha_i T^*_i \leq \frac{\alpha \eta}{2},\;\; \forall \; \mu \in [0,\mu_a'], 
$$ 
and therefore 
\begin{eqnarray*} 
{\dot W}&\leq&(f(T)-f({\bar T}^1))\left(1-\frac{{\bar T}^1}{T} \right)- 
\beta_1 {\bar T^*}_1\left[\frac{{\bar T}^1}{T}+\frac{{\bar T^*}_1V_1T}{T^*_1{\bar V}_1{\bar T}^1}+ 
(1+q_{11}\mu)\frac{{\bar V}_1T^*_1}{V_1{\bar T^*}_1}-3(1+q_{11}\mu)^{1/3}\right]\\ 
&& + 3\beta_1 {\bar T^*}_1(1-(1+q_{11}\mu)^{1/3})-\frac{\alpha \eta}{2} + \alpha V_1. 
\end{eqnarray*} 
Let $\mu_b>0$ be such that for all $\mu \in [0,\mu_b]$, 
$$  
1+q_{11} \mu \in \left[\frac{1}{2},1\right], \quad 3\beta_1 {\bar T^*}_1(1-(1+q_{11}\mu)^{1/3})-\frac{\alpha \eta}{2} 
\leq -\frac{\alpha \eta}{4}. 
$$ 
Let $\mu_1=\min(\mu_a,\mu_a',\mu_b)$ and choose sufficiently large  
$L>0$ so that  
$$3\beta_1 {\bar T^*}_1(1-(1+q_{11}\mu)^{1/3})-\alpha \eta + \alpha V_1<L$$ for all solutions 
of $(\ref{multic4})-(\ref{multic6})$ in $K$ and all $\mu \in [0,\mu_1]$.  
For any $\mu \in [0,\mu_1]$, we have that 
\begin{eqnarray*} 
{\dot W}&\leq &(f(T)-f({\bar T}^1))\left(1-\frac{{\bar T}^1}{T} \right)- 
\beta_1 {\bar T^*}_1\left[\frac{{\bar T}^1}{T}+\frac{{\bar T^*}_1V_1T}{T^*_1{\bar V}_1{\bar T}^1}+ 
(1+q_{11}\mu)\frac{{\bar V}_1T^*_1}{V_1{\bar T^*}_1}-3(1+q_{11}\mu)^{1/3}\right]\\ 
&& -\frac{\alpha \eta}{4} + \alpha V_1,  
\end{eqnarray*} 
where the first two terms are non-positive and $1+q_{11}\mu \in \left[\frac{1}{2},1\right]$. 
Inspecting the first term in $\dot W$,  
we find that there exists $\delta_0>0$ such that  
$$ (f(T)-f({\bar T}^1)\left(1-\frac{{\bar T}^1}{T} \right)<-(L+1)$$ 
for all $T<\delta_0$ and all $\mu \in [0,\mu_1]$. Inspecting the second 
term in $\dot W$, we use Lemma~\ref{minilemma} with
$$ x=(1+q_{11}\mu) \frac{{\bar V}_1T^*_1}{V_1{\bar T^*}_1}, \quad
y= \frac{{\bar T}^1}{T}, \quad
z=1+q_{11}\mu, \quad z_0=\frac{1}{2},$$
and conclude that there exists $\delta_1>0$ such that  
$$ -\beta_1 {\bar T^*}_1\left[\frac{{\bar T}^1}{T}+\frac{{\bar T^*}_1V_1T}{T^*_1{\bar V}_1{\bar T}^1}+ 
(1+q_{11}\mu)\frac{{\bar V}_1T^*_1}{V_1{\bar T^*}_1}-3(1+q_{11}\mu)^{1/3}\right]<-(L+1)$$ 
for all $\frac{T^*_1}{V_1}<\delta_1$ and all $\mu \in [0,\mu_1]$. Finally, 
there exists $\delta_2>0$ such that  
$-\frac{\alpha\eta}{4}+ \alpha V_1 <-\frac{\alpha \eta}{8}$ 
for all $V_1<\delta_2$ and all $\mu \in [0,\mu_1]$. 
Let  
$$\hat K_{\delta}=\{ (T,T^*,V) \in K \cap U | T \geq \delta_0, V_1 \geq \delta_2, 
T^*_1 \geq \delta_1 V_1\}.$$ 
Consider $(T,T^*,V) \in (K \cap U) \backslash \hat K_{\delta}$ 
and let $\mu \in [0,\mu_1]$, then at least one of the following holds: 
\begin{itemize} 
\item $T <\delta_0$, in which case $ \dot W \leq -(L+1) + L \leq -1$; 
\item $T^*_1/V_1 < \delta_1$, in which case $ \dot W \leq -(L+1) + L \leq -1$; 
\item $ V_1 < \delta_2$, in which case $\dot W \leq  -\frac{\alpha \eta}{8}$; 
\end{itemize} 
Hence, for all $(T,T^*,V) \in (K \cap U) \backslash \hat K_{\delta}$ 
and all $\mu \in [0,\mu_1]$, we have 
$$ \dot W \leq -\min(1,\frac{\alpha \eta}{8})<0.$$

The remainder of the proof is the same for both of the above two cases and presented next.  
 
The non-negative function $W(T,T^*,V,\mu)$ is continuous and bounded from above 
on the set $\hat K_{\delta} \times [0,\mu_1]$ because $T,T_1^*,V_1$ are bounded away 
from zero. Hence it attains a finite positive maximum 
$$ w:=\max_{\hat K_{\delta} \times [0,\mu_1]} W(T,T^*,V,\mu)>0.$$ 
Define a new set 
$$ K_{\delta}=\{ (T,T^*,V) \in K \cap U | W(T,T^*,V,\mu) \leq w, \forall  
\mu \in [0,\mu_1]\}.$$ 
By construction, we have that $\hat K_{\delta} \subset K_{\delta} \subset K \cap U$. 
The continuity of $W$ implies that $K_{\delta}$ is closed, and therefore compact in $U$. 
 
It remains to show that all solutions of $(\ref{multic1})-(\ref{multic3})$ in $U$ enter 
and remain in $K_{\delta}$ for all sufficiently large times. Since $K \cap U$ is an  
absorbing set for all $\mu \geq 0$ (Lemma~\ref{bounded}), without loss of 
generality we need to prove this for all solutions in $K \cap U$.  
 
Let $\Phi(t)=(T(t),T^*(t),V(t)) \in K \cap U$ be a solution of $(\ref{multic1})-(\ref{multic3})$ 
for some fixed $\mu \in [0,\mu_1]$. Observe that in the set $(K \cap U) \backslash \hat K_{\delta}$,  
the inequality $\dot W \leq -\min(1,\frac{\alpha \eta}{8})<0$ holds. Since $W \geq 0$, there 
exists $t_0 \geq 0$ such that $\Phi(t_0) \in \hat K_{\delta} \subset K_{\delta}$. We will 
show that $\Phi(t) \in K_{\delta}$ for all $t \geq t_0$. For the sake of contradiction, 
let us suppose that there exists $t_1>t_0$ such that $\Phi(t_1) \notin K_{\delta}$. 
Then there exists $t_2 \in [t_0,t_1)$ such that $\Phi(t_2) \in K_{\delta}$ and 
$\Phi(t) \notin K_{\delta}$ for all $t\in(t_2,t_1]$. On the one hand, we have 
that  
$$ W(\Phi(t_2),\mu) \leq w < W(\Phi(t_1),\mu)$$ 
by definition of $K_{\delta}$. On the other hand, for all  
$t\in(t_2,t_1]$, we have $\Phi(t) \notin K_{\delta}$ and consequently 
$\Phi(t) \notin \hat K_{\delta}$ so that $\frac{d}{dt} W(\Phi(t),\mu)=\dot W <0$. 
This contradiction shows that $\Phi(t) \in K_{\delta}$ for all $t \geq t_0$ and 
concludes the proof of the Theorem. 
\end{proof}

\begin{stel}\label{global} 
 
Let the assumptions of Lemma~\ref{invertible} hold, let $U$ be the set from Theorem $\ref{compactset}$, and define  
 
$$ U'=\{ (T,T^*,V) \in \reals^{2n+1}_+ | \; \; T^*_1+V_1>0 \}\supset U.$$ 
 
Then there exist $\mu_0>0$ and a continuous map $E: [0,\mu_0] \rightarrow U$  
such that  
\begin{enumerate} 
 
\item 
 $E(0)=E_1$ (where $E_1$ is the same as in Lemma~\ref{invertible}),  
and  
$E(\mu)$ is an equilibrium of $(\ref{multic1})-(\ref{multic3})$ or of $(\ref{multic4})-(\ref{multic6})$ for  
all $\mu \in [0,\mu_0]$; 
\item For each $\mu \in [0,\mu_0]$,  $E(\mu)$ is  
a globally asymptotically stable equilibrium  of $(\ref{multic1})-(\ref{multic3})$ or  
of $(\ref{multic4})-(\ref{multic6})$ in $U'$. 
 
\end{enumerate} 
 
\end{stel} 
 
\begin{proof} 
To prove the first assertion, we begin by noting that for $\mu=0$, $E_1$ is a stable hyperbolic equilibrium 
of $(\ref{multic1})-(\ref{multic3})$ or of $(\ref{multic4})-(\ref{multic6})$ by Lemma $\ref{invertible}$.  
Since the vector field of $(\ref{multic1})-(\ref{multic3})$ and  $(\ref{multic4})-(\ref{multic6})$ is  
linear in $\mu$, by  the Implicit Function Theorem there exist $h>0$ and a continuous map  
$E:(-h,h) \rightarrow \reals^{2n+1}$ such that $E(\mu)$ is an equilibrium of  
$(\ref{multic1})-(\ref{multic3})$ or $(\ref{multic4})-(\ref{multic6})$  
for all $\mu \in (-h,h)$. The fact that $E(\mu) \in U$ for all $\mu \in [0,h)$ follows from 
Proposition~\ref{reduce} and the fact that $\bar T_1 < \bar T_i,\ i \geq 2$. 
Note that for $\mu>0$, $E(\mu)$ may be positive (if $Q$ is irreducible) or  
non-negative (if $Q$ is reducible). Nevertheless, in  both cases, $\mu>0$ implies $E(\mu) \in U$. 
 
The proof of the second assertion is based on the result of Smith and Waltman (Corollary 2.3 
in \cite{smith-waltman}).  We have already established the fact that $E(0)$ is a stable hyperbolic 
equilibrium of $(\ref{multic1})-(\ref{multic3})$ or $(\ref{multic4})-(\ref{multic6})$.  
By Theorem~\ref{multistrain}, $E(0)$ is globally 
asymptotically stable in $U'$ for $\mu=0$. In addition, by Theorem~\ref{compactset} 
there exist $\mu_0>0$ and a compact set $K_{\delta} \subset U$ such that for each 
$\mu \in[0,\mu_0]$, and each solution $(T(t),T^*(t),V(t))$ of $(\ref{multic1})-(\ref{multic3})$ 
or $(\ref{multic4})-(\ref{multic6})$ 
in $U$, there exists $t_0>0$ such that $(T(t),T^*(t),V(t)) \in K_{\delta}$ for all $t>t_0$. 
Hence, the condition (H1) of Corollary 2.3 in \cite{smith-waltman} holds. The 
Proposition 2.3 itself then implies the global stability of $E(\mu)$ in $U$ for all 
sufficiently small $\mu\geq 0$. Finally, solutions of $(\ref{multic1})-(\ref{multic3})$ 
or $(\ref{multic4})-(\ref{multic6})$ starting in $U'$ enter $U$ instantaneously, hence global  
stability of $E(\mu)$ in $U'$ follows as well. 
\end{proof} 
 
\section*{Appendix: Inclusion of loss of virus in the model} 
 
\subsection*{Single-strain} 
When taking the loss of the virus particle upon infection into account, model $(\ref{hiv1})$ becomes 
\begin{eqnarray}\label{hiv1a} 
{\dot T}&=&f(T)-kVT\nonumber \\ 
{\dot T^*}&=&kVT-\beta T^*\nonumber \\ 
{\dot V}&=&N\beta T^*-\gamma V-kVT, 
\end{eqnarray} 
We still assume that the growth rate of the healthy cell population is given by $(\ref{T0})$, hence 
$E_0=(T_0,0,0)$ is still an equilibrium of $(\ref{hiv1a})$.  
A second, positive equilibrium may exist if the following quantities are positive: 
\begin{equation}\label{eqa} 
{\bar T}=\frac{\gamma}{k(N-1)},\;\; {\bar T^*}=\frac{f({\bar T})}{\beta},\;\; {\bar V}=\frac{f({\bar T})}{k{\bar T}}. 
\end{equation} 
Note that this is the case iff $N>1$ and $f\left(\frac{\gamma}{k(N-1)}\right)>0$, or equivalently by $(\ref{T0})$  
that ${\bar T}=\frac{\gamma}{k(N-1)}<T_0$. In terms of the basic reproduction number 
$$ 
{\cal R}^0:=\frac{k(N-1)}{\gamma}T_0 =\frac{T_0}{\bar T}, 
$$ 
existence of a positive equilibrium is therefore equivalent to ${\cal R}^0>1$.
Assuming that ${\cal R}^0>1$, we will still denote this disease steady state by  
$E=({\bar T},{\bar T^*},{\bar V})$. We introduce the following condition. 
$$ 
{\bf (C')}\;\;f'(c)+\frac{k}{\gamma}f({\bar T})\leq 0, \textrm{ for all } c\in [0,T_0]. 
$$ 
Note that this condition is satisfied when $f(T)$ is a decreasing function with sufficiently large 
negative derivative. 
 
\begin{stel}\label{1straina} 
Let ${\bf (C')}$ hold.   
Then the equilibrium $E$ is globally asymptotically stable for $(\ref{hiv1a})$  
with respect to initial conditions satisfying $T^*(0)+V(0)>0$. 
\end{stel} 
 
\begin{proof} 
Consider the following function on $\interior(\reals^3_+)$: 
$$ 
W=(N-1)\int_{{\bar T}}^T \left(1-\frac{{\bar T}}{\tau}\right) d \tau  
+N\int_{{\bar T^*}}^{T^*} \left(1-\frac{{\bar T^*}}{\tau}\right)d \tau+ 
\int_{{\bar V}}^V \left(1-\frac{{\bar V}}{\tau}\right) d \tau. 
$$ 
Then  
\begin{eqnarray*} 
{\dot W}&=&(N-1)(f(T)-kVT)\left(1-\frac{{\bar T}}{T} \right)+N(kVT-\beta T^*)\left(1-\frac{{\bar T^*}}{T^*} \right)+ 
(N\beta T^*-\gamma V-kVT)\left(1-\frac{{\bar V}}{V} \right)\\ 
&=&(N-1)f(T)\left(1-\frac{{\bar T}}{T}\right)-NkVT\frac{{\bar T^*}}{T^*}+N\beta{\bar T^*}-N\beta T^* \frac{{\bar V}}{V}+\gamma {\bar V} 
+k{\bar V}T\\ 
&=&(N-1)(f(T)-f({\bar T}))\left(1-\frac{{\bar T}}{T}\right)+(N-1)f({\bar T})\left(1-\frac{{\bar T}}{T}\right) 
+N\beta{\bar T^*}\left[2-\frac{VT{\bar T^*}}{{\bar V}{\bar T}T^*}-\frac{T^*{\bar V}}{{\bar T^*}V}\right] 
-\beta{\bar T^*}+\beta {\bar T^*}\frac{T}{{\bar T}}\\ 
&=&(N-1)(f(T)-f({\bar T}))\left(1-\frac{{\bar T}}{T}\right)+(N-1)\beta {\bar T^*}\left(1-\frac{{\bar T}}{T}\right) 
+N\beta{\bar T^*}\left[2-\frac{VT{\bar T^*}}{{\bar V}{\bar T}T^*}-\frac{T^*{\bar V}}{{\bar T^*}V}\right] 
-\beta{\bar T^*}+\beta {\bar T^*}\frac{T}{{\bar T}}\\ 
&=&(N-1)(f(T)-f({\bar T}))\left(1-\frac{{\bar T}}{T}\right)+\beta{\bar T^*}\left(-2+\frac{{\bar T}}{T}+\frac{T}{{\bar T}}\right) 
+N\beta{\bar T^*}\left[3-\frac{VT{\bar T^*}}{{\bar V}{\bar T}T^*}-\frac{T^*{\bar V}}{{\bar T^*}V}-\frac{{\bar T}}{T}\right]\\ 
&=&\left[(N-1)(f(T)-f({\bar T})){\bar T}+\beta {\bar T^*}(T-{\bar T}) \right]\frac{(T-{\bar T})}{T{\bar T}} 
+N\beta{\bar T^*}\left[3-\frac{VT{\bar T^*}}{{\bar V}{\bar T}T^*}-\frac{T^*{\bar V}}{{\bar T^*}V}-\frac{{\bar T}}{T}\right]\\ 
\end{eqnarray*} 
where we used $(\ref{eqa})$ repeatedly; in particular in the second, third and fourth equation.  
By the mean value theorem there is some $c\in (T,{\bar T})$ or $({\bar T},T)$ such that  
$$ 
f(T)-f({\bar T})=f'(c)(T-{\bar T}), 
$$ 
hence using $(\ref{eqa})$ once more 
\begin{equation*} 
{\dot W}=(N-1)\left[f'(c)+\frac{k}{\gamma}f({\bar T}) \right]\frac{(T-{\bar T})^2}{T} 
+N\beta{\bar T^*}\left[3-\frac{VT{\bar T^*}}{{\bar V}{\bar T}T^*}-\frac{T^*{\bar V}}{{\bar T^*}V}-\frac{{\bar T}}{T}\right]. 
\end{equation*} 
The first term is non-positive by ${\bf (C')}$ and because we can assume that $T\leq T_0$ by dissipativity  
(see Lemma $\ref{boundedKVT}$ later).  
The second term is non-positive as well since the geometric mean of $3$  
non-negative numbers is not larger than the arithmetic mean of those numbers.  
We conclude that ${\dot W}\leq 0$ in $\interior(\reals^3_+)$, hence local stability of $E$ follows.  
Notice that ${\dot W}$ equals zero if and only if both the first term and the second term are zero,  
This happens at points where: 
$$ 
\frac{{\bar T}}{T}=1\textrm{ and } \frac{{\bar T^*}V}{T^*{\bar V}}=1. 
$$ 
Then LaSalle's Invariance Principle implies that all bounded solutions (and as before, solutions are easily shown  
to be bounded, see also Lemma~\ref{boundedKVT} later) in $\interior(\reals^3_+)$ converge to the largest invariant set in  
$$ 
M=\{(T,T^*,V)\in \interior(\reals^3_+)\;|\; \frac{{\bar T}}{T}=1,\;\; \frac{{\bar T^*}V}{T^*{\bar V}}=1\}. 
$$ 
It is clear that the largest invariant set in $M$ is the singleton $\{E\}$.  
Finally, note that forward solutions starting on  
the boundary of $\reals^3_+$ with either $T_1(0)$ or $V_1(0)$ positive,  
enter $\interior(\reals^3_+)$ instantaneously. This concludes the proof. 
\end{proof}

\subsection*{Competitive exclusion} 
Now we modify the multi-strain model $(\ref{multi1})-(\ref{multi3})$ to 
\begin{eqnarray} 
{\dot T}&=&f(T)-kVT, \quad T \in \reals_+\label{hiv2a} \\ 
{\dot T^*}&=&KVT-B T^*,\quad T^* \in \reals^n_+\label{hiv2b} \\ 
{\dot V}&=&\hat N B T^*-\Gamma V-KVT, \quad V \in \reals^n_+\label{hiv2c}, 
\end{eqnarray} 
where $k=(k_1,...,k_n)$, $K=\diag(k_1,...,k_n)$, $B=\diag(\beta_1,...,\beta_n)$, 
$\hat N=\diag(N_1,...,N_n)$, and $\Gamma=\diag(\gamma_1,...,\gamma_n)$. Suppose 
that each strain is capable to persist at steady state by itself, that is, $N_i>1$ and 
$\bar T_i =\frac{\gamma_i}{k_i(N_i-1)} < T_0$ and denote the corresponding equilibria also  
by $E_1,\dots, E_n$. Assume that   
\begin{equation}\label{nogeens} 
0< \bar T_1 \leq \bar T_2 \leq \ldots \leq \bar T_n < T_0. 
\end{equation} 
In addition, suppose that ${\bf (C')}$ holds with $\bar T=\bar T_1$. Then we have the following. 
\begin{stel}\label{multistrain2}   
The single strain equilibrium $E_1$ is globally asymptotically stable for $(\ref{hiv2a})-(\ref{hiv2c})$  
with respect to initial conditions satisfying $T_1^*(0)+V_1(0)>0$. 
\end{stel} 
\begin{proof} 
Consider the function $W$ defined on $U:=\{(T,T^*,V)\in \reals^{2n+1}\;|\; T,T_1^*,V_1>0\}$ as 
\begin{eqnarray*} 
 W &=& (N_1-1)\int_{{\bar T_1}}^T \left(1-\frac{{\bar T_1}}{\tau}\right) d \tau  
+N_1\int_{{\bar T_1^*}}^{T_1^*} \left(1-\frac{{\bar T_1^*}}{\tau}\right)d \tau+ 
\int_{{\bar V_1}}^{V_1} \left(1-\frac{{\bar V_1}}{\tau}\right) d \tau\\ 
&& + \sum_{i=2}^n \frac{N_1-1}{N_i-1}(N_i T^*_i +V_i). \end{eqnarray*} 
Computing $\dot W$, we find that 
\begin{eqnarray*} 
 \dot W &=& (N_1-1)\left[f'(c)+\frac{k}{\gamma}f({\bar T_1}) \right]\frac{(T-{\bar T_1})^2}{T} 
+N_1\beta_1{\bar T_1^*}\left[3-\frac{V_1 T{\bar T_1^*}} 
{{\bar V_1}{\bar T_1}T_1^*}-\frac{T_1^*{\bar V_1}}{{\bar T_1^*}V_1}-\frac{{\bar T_1}}{T}\right]\\ 
 & & +\sum_{i=2}^n \left(-k_i V_i (T-\bar T_1) + \frac{(N_1-1)}{N_i-1} 
(N_i k_i V_i T-N_i \beta_i T^*_i  + N_i \beta_i T^*_i - \gamma_i - k_i V_i T)  \right). 
\end{eqnarray*} 
After simplifications, we have 
\begin{eqnarray*} 
 \dot W &=& (N_1-1)\left[f'(c)+\frac{k}{\gamma}f({\bar T_1}) \right]\frac{(T-{\bar T_1})^2}{T} 
+N_1\beta_1{\bar T_1^*}\left[3-\frac{V_1 T{\bar T_1^*}} 
{{\bar V_1}{\bar T_1}T_1^*}-\frac{T_1^*{\bar V_1}}{{\bar T_1^*}V_1}-\frac{{\bar T_1}}{T}\right]\\ 
 & & -(N_1-1) \sum_{i=2}^n k_i V_i (\bar T_i -\bar T_1). 
\end{eqnarray*} 
The first term is non-positive since ${\bf (C')}$ with ${\bar T}={\bar T_1}$ holds and because  
$T\leq T_0$ by disspiativity (see Lemma $\ref{boundedKVT}$ later). The second term is non-positive is well,  
and so is the third by $(\ref{nogeens})$. Thus ${\dot W}\leq 0$ which already implies that $E_1$ is stable.  
An application of LaSalle's Invariance Principle shows that all bounded solutions in $U$  
(boundedness follows from Lemma~$\ref{boundedKVT}$ which is proved later)  
converge to the largest invariant set in  
$$ 
\left\{(T,T^*_1,\dots,T^*_n,V_1,\dots,V_n)\in U\;|\;\frac{{\bar T^1}}{T}=1,\;\; \frac{{\bar T^*_1}V_1}{T^*_1{\bar V_1}}=1 
,\;\;V_i=0,\;\; i>2 \right\}, 
$$ 
which is easily shown to be the singleton $\{E_1\}$. Finally, solutions on the boundary of $U$  
with $T_1^*(0)+V_1(0)>0$ enter $U$ instantaneously, which concludes the proof. 
\end{proof} 
\subsection*{Adding mutations} 
We modify the model $(\ref{hiv2a})-(\ref{hiv2c})$ to account for mutations. Again, we consider two 
alternative models 
\begin{eqnarray}\label{hiv3a} 
{\dot T}&=&f(T)-kVT,\quad T \in \reals_+\nonumber \\ 
{\dot T^*}&=&P(\mu)KVT- 
B T^*, \quad T^* \in \reals^n_+\nonumber \\ 
{\dot V}&=&\hat N B T^*-\Gamma V-KVT, \quad V \in \reals^n_+, 
\end{eqnarray} 
and 
\begin{eqnarray}\label{hiv3b} 
{\dot T}&=&f(T)-kVT,\quad T \in \reals_+\nonumber \\ 
{\dot T^*}&=&KVT- 
B T^*, \quad T^* \in \reals^n_+\nonumber \\ 
{\dot V}&=&P(\mu)\hat N B T^*-\Gamma V-KVT,\quad V \in \reals^n_+,  
\end{eqnarray} 
where $k,K,B,\hat N, \Gamma$ are the same as before, and $P(\mu)=I +\mu Q$ and $Q$ is a stochastic matrix with non-negative 
off-diagonal entries. 
\begin{lemma}\label{boundedKVT} 
Both systems $(\ref{hiv3a})$ and $(\ref{hiv3b})$ are dissipative, i.e.  
there is some compact set $K$ such that every solution eventually enters $K$ and remains in $K$  
forever after. 
\end{lemma} 
 
\begin{proof} 
The proof is similar to the proof of Lemma~\ref{bounded} and will be omitted. 
\end{proof}

\begin{lemma}\label{invertibleKVT} 
For $\mu =0$, let all single strain equilibria $E_1,E_2\dots,E_n$ exist  
for either $(\ref{hiv3a})$ or $(\ref{hiv3b})$, and assume that 
\begin{equation}\label{c1KVT} 
{\bar T^1}<{\bar T^2}<\dots<{\bar T^n}<{\bar T^{n+1}}:=T_0, 
\end{equation} 
and  
\begin{equation}\label{c2KVT} 
f'({\bar T}^j)\leq 0,\textrm{ for all }j=1,\dots, n+1. 
\end{equation} 
Then the Jacobian matrices of $(\ref{hiv3a})$ or $(\ref{hiv3b})$,  
evaluated at any of the $E_i$'s, $i=1,\dots, n+1$ (where $E_{n+1}:=E_0$) have the following properties: 
$J(E_i)$ has $i-1$ eigenvalues (counting multiplicities) in the open right half plane and  
$2(n+1)-i$ eigenvalues in the open left half plane. In particular, $J(E_1)$ is Hurwitz.  
\end{lemma} 
\begin{proof} 
The proof is similar to that of Lemma $\ref{invertible}$.  
The only difference is that the entries of the Jacobian matrices change. In particular,  
the $(3,1)$ and $(3,3)$ entry of $A_1^i$ now become $-k_i{\bar V_i}$ and $-\gamma_i-k{\bar T^i}$ respectively,  
but by $(\ref{c1KVT})$ and Lemma $3.4$ in \cite{hiv}, $A_1^i$ is still Hurwitz. 
 
\end{proof}

To study equilibria of systems $(\ref{hiv3a})$ and $(\ref{hiv3b})$, we introduce 
the matrix 
\begin{equation} 
A(\mu) =\Gamma^{-1}(\hat N P(\mu) - I)K, 
\label{AKVT}\end{equation} 
which has non-negative off-diagonal entries for $\mu>0$ and  
$$A(0)=\diag\left(\frac{k_1(N_1-1)}{\gamma_1},\ldots, \frac{k_n(N_n-1)}{\gamma_n} \right)= 
\diag\left(\frac{1}{\bar T_1},\ldots, \frac{1}{\bar T_n} \right).$$ 
 
Clearly, Proposition~\ref{eigenvectors} holds with $A(\mu)$ given by (\ref{AKVT}). 
Hence, we have the following.

\begin{prop}\label{reduceKVT} 
Let the assumptions of Lemma $\ref{invertibleKVT}$ hold and suppose that  
the strains are renumbered so that $A(\mu)$ has the form (\ref{Amu}).  
Let $E_j(\mu)=(\hat T_j(\mu),\hat T^*_j(\mu),\hat V_j(\mu))$ denote the nontrivial equilibria of both  
$(\ref{hiv3a})$ and $(\ref{hiv3b})$ 
for small $\mu>0$. Then  
\begin{enumerate} 
\item $E_j(\mu)$ is positive if and only if $\frac{1}{\hat T_j(\mu)}$ is  
an eigenvalue of $A(\mu)$ with a positive eigenvector $V_j$. 
\item $E_j(\mu)$ is non-negative if and only if $\frac{1}{\hat T_j(\mu)}$ is 
an eigenvalue of $A(\mu)$ with a non-negative eigenvector $V_j$. 
\item $E_j(\mu) \notin \reals^{2n+1}_+$ if and only if  $\frac{1}{\hat T_j(\mu)}$ is an eigenvalue of  
$A(\mu)$ with eigenvector $V_j$ which is not sign-definite. 
\end{enumerate} 
\end{prop}

\begin{proof} 
We will prove the Proposition only for system $(\ref{hiv3a})$ (the proof for  
$(\ref{hiv3b})$ is similar). Observe that at 
 equilibrium,  
$\hat T^*_j(\mu)=({\hat N}B)^{-1}(\Gamma+K \hat T_j(\mu)) {\hat V_j}(\mu)$. Hence, the the signs of the corresponding components of $\hat T^*_j(\mu)$ and ${\hat V_j}(\mu)$ 
are the same.  
Substituting the above expression into $(\ref{hiv3a})$,  
we find that ${\hat V_j}(\mu)$ must satisfy 
$$ 
[\Gamma^{-1} ({\hat N}P(\mu)-I)K-\frac{1}{{\hat T_j}(\mu)}I]{\hat V_j}(\mu)= 
[A(\mu)-\frac{1}{{\hat T_j}(\mu)}I]{\hat V_j}(\mu)=0. 
$$ 
Thus for each nontrivial equilibrium $E_j(\mu)$, the quantity $\frac{1}{{\hat T_j}(\mu)}$ 
must be an eigenvalue of $A(\mu)$ and ${\hat V_j}(\mu)$ must be a multiple of the 
corresponding eigenvector $V_j$. If $V_j$ is not sign definite, it follows that  
$E_j(\mu) \notin \reals^{2n+1}_+$. 
For all $V_j \geq 0$, the components of $E_j(\mu)$ are uniquely determined via 
$$ \hat V_j(\mu)=\frac{f(\hat T_j(\mu))}{\hat T_j(\mu) k' V_j} V_j, \quad 
\hat T^*_j(\mu)=({\hat N}B)^{-1}(\Gamma+ K \hat T_j(\mu)) {\hat V_j}(\mu).$$ 
Hence $E_j(\mu)$ is positive (non-negative) if and only if $V_j$ is positive (non-negative). 
\end{proof}  
 
\subsection*{Lower bounds}

\begin{lemma}\label{lowerboundKVT} 
Suppose that $(\ref{c1KVT})$ holds. Then 
there exist $\eta,\mu_0>0$ such that 
$$\liminf_{t\to\infty} 1' V(t) \geq \eta >0$$for any $\mu\in[0,\mu_0]$ and for any solution of $(\ref{hiv3a})$ and $(\ref{hiv3b})$ with $1' V(t)>0$. 
\end{lemma}

\begin{proof} 
We will prove the claim for system $(\ref{hiv3a})$(the proof for $(\ref{hiv3b})$ is similar). The proof consists of two parts.  
We first show that there exist  $\eta_0,\mu_0>0$ such 
 that $\liminf_{t\to\infty} 1'(T^*(t)+V(t)) \geq \eta_0 >0$ for all solutions with $T^*(t),V(t) \not=0$. 
 We choose $n$ positive numbers $\tilde N_i$ so that  
$\frac{\gamma_i +k_i T_0}{k_i T_0} < \tilde N_i<N_i$  
for all $1 \leq i \leq n$.  
This is possible since we assume  
$\bar T_i =\frac{\gamma_i}{k_i (N_i-1)} < T_0$ which is equivalent to 
$N_i > \frac{\gamma_i +k_i T_0}{k_i T_0}$. Let $v=(\tilde N,1)$. It follows that 
$$ v'\begin{pmatrix} -B & K T_0\cr \hat N B & -\Gamma-KT_0\cr\end{pmatrix}= 
\biggl(b_1(N_1-\tilde N_1),...,b_n(N_n-\tilde N_n), k_1 T_0 \tilde N_1 - (\gamma_1+k_1 T_0),..., 
k_n T_0 \tilde N_n - (\gamma_n+k_n T_0)\biggr)$$ 
is a positive vector. By continuity, there exists a $\mu_0>0$ such that 
$$v'M(T_0,\mu),\quad {\rm where}\  
M(T,\mu):=\begin{pmatrix} -B & P(\mu) K T\cr \hat N B & -\Gamma-K T\cr\end{pmatrix}$$ 
is a positive vector for all $\mu \in [0,\mu_0]$.  
 
Consider a system  
\begin{eqnarray} 
{\dot T}&=&f(T)-k'VT,\ T \in \reals_+\label{persKVT1}\\ 
{\dot T^*}&=&P(\mu)KVT-B T^*,\ T^* \in \reals^n_+,\label{persKVT2}\\ 
{\dot V}&=&{\hat N}B T^*-\Gamma V -KVT,\ V\in \reals^n_+,\label{persKVT3}\\ 
{\dot \mu}&=& 0,\ \mu \in [0,\mu_0].\label{persKVT4} 
\end{eqnarray} 
Let $K'$ be the forward invariant compact set for  
$(\ref{hiv3a})$  
established in Lemma~\ref{boundedKVT} and define $K=K'\times[0,\mu_0]$. It is clear that 
$K$ is compact and forward invariant under $(\ref{persKVT1})-(\ref{persKVT4})$ 
The set $K_0=([0,T_0]\times 0 \times 0 \times [0,\mu_0]) \cap K$ is 
clearly a compact forward invariant subset of $K$.  
 
Let $\Pi(T^*,V):=v'(T^*,V)$. The function $\Pi$ is clearly 
smooth, zero on $K_0$, and positive on $K\backslash K_0$. Furthermore, 
$$ \frac{\dot \Pi}{\Pi}= \psi := 
\frac{v'M(T,\mu)(T^*,V)}{v'(T^*,V)}$$ 
is lower semi-continuous on $K$ by Lemma~\ref{liminf} once we define the value of $\psi$ on $K_0$ 
as 
$$ \psi(T,\mu)=\min_{i=1,...,n} 
\frac{v'M(T,\mu)_i}{v_i} .$$ 
We note that the function $\psi(T,\mu)$ is continuous in $(T,\mu)$. 
Since all solutions of $(\ref{persKVT1})-(\ref{persKVT4})$ in $K_0$ have the property that  
$\lim_{t\to\infty}T(t)=T_0$, it implies that $\psi(T(t),\mu)>0$ for all sufficiently large $t$. 
Hence by Theorem~\ref{patrickdoeshofbauer}, the set $K_0$ is a uniform strong repellor in $K$. 
If we use the $L^1$-norm of $(T^*,V)$ as the distance function to $K_0$, we find that there 
exists an $\eta_0>0$ such that 
$$ \liminf_{t\to\infty} 1'(T^*+V) \geq \eta_0$$ 
for all solutions of $(\ref{persKVT1})-(\ref{persKVT4})$ in $K\backslash K_0$.  
 
To complete the proof, we need to show that there exists $\eta>0$ such that 
$\liminf_{t\to\infty} 1'V(t) \geq \eta >0$ for all solutions with $1'V(t)>0$.  
Observe that $1'V(t)>0$ implies that $1'T^*(t)>0$. Hence by the result of  
part one, we have that $\liminf_{t\to\infty} 1'(T^*(t)+V(t)) \geq \eta_0 >0$, or  
equivalently, $1'T^*(t) > \eta_0/2 - 1'V(t)$ for all sufficiently large $t$.  
From  (\ref{persKVT3}), we have that 
$$ 1'\dot V \geq \sum_{i=1}^n N_i \beta_i T^*_i - \sum_{i=1}^n (\gamma_i + k_i T) V_i 
\geq \sum_{i=1}^n N_i \beta_i T^*_i - \sum_{i=1}^n (\gamma_i + k_i T_0) V_i.$$Hence, 
$$ 1'\dot V \geq A_0 \biggl(\frac{\eta_0}{2} - 1'V(t)\biggr)-A_1 1'V(t), \quad A_0:=\min_{i} (N_i \beta_i)>0, 
\ A_1:=\max_{i} (\gamma_i+k_i T_0) >0$$ 
holds for large $t$. It follows immediately that 
$$ \liminf_{t\to\infty} 1'V(t) \geq \eta=\frac{\eta_0 A_0}{2(A_0+A_1)}>0.$$ 
\end{proof}

\subsection*{Existence of absorbing compact set for small $\mu>0$.} 
 
\begin{stel}\label{compactset-KVT} 
 Let $K$ be the absorbing compact set established  
in Lemma~\ref{boundedKVT}, and let  
$$ U=\{ (T,T^*,V) \in \reals^{2n+1}_+ | T,T^*_1,V_1>0 \}.$$ 
 Suppose that there exists $\epsilon>0$  such that

$$ 
{\bf (C_{\epsilon})}\;\;f'(c)+\frac{k_1}{\gamma_1}f({\bar T_1})\leq -\epsilon <0,  
\textrm{ for all } c\in [0,T_0]. 
$$ 
Then there exist $\mu_1>0$ and a compact set $K_{\delta} \subset U$ such that for  
any $\mu\in[0,\mu_1]$ and for any solution of system $(\ref{hiv3a})$ in  
$U$, 
there exists a $t_0>0$ such that  
$ (T(t),T^*(t),V(t)) \in K_{\delta}$ for  
all $t >t_0$. \\ An identical statement holds for system $(\ref{hiv3b})$.  
  
\end{stel} 
 
\begin{proof} 
\noindent (a) 
We first prove the statement for system $(\ref{hiv3a})$. Consider the function 
\begin{eqnarray*} 
 W &=& (N_1-1)\int_{{\bar T_1}}^T \left(1-\frac{{\bar T_1}}{\tau}\right) d \tau  
+N_1\int_{{\bar T_1^*}}^{T_1^*} \left(1-\frac{{\bar T_1^*}}{\tau}\right)d \tau+ 
\int_{{\bar V_1}}^{V_1} \left(1-\frac{{\bar V_1}}{\tau}\right) d \tau\\ 
&& + \sum_{i=2}^n \frac{N_1-1}{N_i-1}(N_i T^*_i +V_i). \end{eqnarray*} 
Computing $\dot W$ for the system $(\ref{hiv3a})$, we obtain 
\begin{eqnarray*} 
 \dot W &=& (N_1-1)\left[f'(c)+\frac{k}{\gamma}f({\bar T_1}) \right]\frac{(T-{\bar T_1})^2}{T} 
+N_1\beta_1{\bar T_1^*}\left[3-\frac{V_1 T{\bar T_1^*}} 
{{\bar V_1}{\bar T_1}T_1^*}-\frac{T_1^*{\bar V_1}}{{\bar T_1^*}V_1}-\frac{{\bar T_1}}{T}\right]\\ 
 & & -(N_1-1) \sum_{i=2}^n k_i V_i (\bar T_i -\bar T_1) +  
\mu N_1 \frac{T^*_1-\bar T^*_1}{T^*_1} \sum_{j=1}^n q_{1j} k_j V_j T 
+\mu (N_1-1) \sum_{i=2}^n \frac{N_i}{N_i-1} \sum_{j=1}^n q_{ij} k_j V_j T, 
\end{eqnarray*} 
Recombining the terms, we find that 
\begin{eqnarray*} 
\dot W &=& (N_1-1)\left[f'(c)+\frac{k}{\gamma}f({\bar T_1}) \right]\frac{(T-{\bar T_1})^2}{T} 
+N_1\beta_1{\bar T_1^*}\left[3-\frac{V_1 T{\bar T_1^*}} 
{{\bar V_1}{\bar T_1}T_1^*}-\frac{T_1^*{\bar V_1}}{{\bar T_1^*}V_1}-\frac{{\bar T_1}}{T}\right]\\ 
 & & -(N_1-1) \sum_{i=2}^n k_i V_i (\bar T_i -\bar T_1)  
+\mu (N_1-1) \sum_{i=1}^n \frac{N_i}{N_i-1} \sum_{j=1}^n q_{ij} k_j V_j T \\ && 
-\mu N_1 q_{11} \frac{\bar T^*_1 V_1 T}{T^*_1} -  
\mu N_1 \frac{\bar T^*_1}{T^*_1} \sum_{j=1}^n q_{1j} k_j V_j T, 
\end{eqnarray*} 
where the last term is clearly non-positive. Let  
\begin{eqnarray*} 
 \alpha &= &(N_1-1) \min_{i\geq 2} k_i (\bar T_i -\bar T_1) >0,\\ 
 L & =& \sup_{K} (N_1-1) \sum_{i=1}^n \frac{N_i}{N_i-1} \sum_{j=1}^n q_{ij} k_j V_j T \geq 0. 
\end{eqnarray*} 
By Lemma~\ref{lowerboundKVT}, 
 there exist $\eta,\mu_a>0$ such that $1'V(t) > \eta$  
for all $\mu \in [0,\mu_a]$ 
 and all sufficiently large $t$. Hence, by shifting 
time forward if necessary, we have the inequality 
\begin{eqnarray*} 
\dot W &\leq & -\epsilon (N_1-1)\frac{(T-{\bar T_1})^2}{T}+N_1\beta_1{\bar T_1^*}\left[3-\frac{V_1 T{\bar T_1^*}} 
{{\bar V_1}{\bar T_1}T_1^*}-\frac{T_1^*{\bar V_1}}{{\bar T_1^*}V_1}-\frac{{\bar T_1}}{T}\right]\\ 
 & & -\alpha (\eta -V_1) +\mu L -\mu N_1 q_{11} \frac{\bar T^*_1 V_1 T}{T^*_1}, 
\end{eqnarray*} 
which holds in $K$ for all $\mu \in [0,\mu_a]$. We combine the second and the last 
terms to obtain 
\begin{eqnarray*} 
\dot W &\leq & -\epsilon (N_1-1)\frac{(T-{\bar T_1})^2}{T}+N_1\beta_1{\bar T_1^*}\left[3- 
(1+q_{11}\mu)\frac{V_1 T{\bar T_1^*}} 
{{\bar V_1}{\bar T_1}T_1^*}-\frac{T_1^*{\bar V_1}}{{\bar T_1^*}V_1}- 
\frac{{\bar T_1}}{T}\right]\\ 
 & & -\alpha (\eta -V_1) +\mu L. 
\end{eqnarray*} 
Further, we rewrite the above inequality as 
\begin{eqnarray*} 
\dot W &\leq & -\epsilon (N_1-1)\frac{(T-{\bar T_1})^2}{T}+N_1\beta_1{\bar T_1^*} 
\left[3(1+q_{11}\mu)^{1/3}- 
(1+q_{11}\mu)\frac{V_1 T{\bar T_1^*}} 
{{\bar V_1}{\bar T_1}T_1^*}-\frac{T_1^*{\bar V_1}}{{\bar T_1^*}V_1}- 
\frac{{\bar T_1}}{T}\right]\\ 
 & & -\alpha (\eta -V_1) +\mu L + 3 N_1\beta_1{\bar T_1^*}\left[1-(1+q_{11}\mu)^{1/3}\right]. 
\end{eqnarray*} 
Let $\mu_b>0$ be such that for all $\mu \in [0,\mu_b]$, 
$$ (1+q_{11}\mu) \in [\frac{1}{2},1], \quad 
-\alpha \eta +\mu L + 3 N_1\beta_1{\bar T_1^*}\left[1-(1+q_{11}\mu)^{1/3}\right]  
\leq -\frac{\alpha \eta}{2} .$$ 
Now we let $\mu_1=\min[\mu_a,\mu_b]$, so that for all $\mu \in [0,\mu_1]$ 
and all points in $K$, 
\begin{eqnarray*} 
\dot W &\leq & -\epsilon (N_1-1)\frac{(T-{\bar T_1})^2}{T}+N_1\beta_1{\bar T_1^*} 
\left[3(1+q_{11}\mu)^{1/3}- 
(1+q_{11}\mu)\frac{V_1 T{\bar T_1^*}} 
{{\bar V_1}{\bar T_1}T_1^*}-\frac{T_1^*{\bar V_1}}{{\bar T_1^*}V_1}- 
\frac{{\bar T_1}}{T}\right]\\ 
 & & -\frac{\alpha \eta}{2} +\alpha V_1. 
\end{eqnarray*} 
Let $L_1=\alpha \sup_{K} V_1$. 
Inspecting the first term in $\dot W$,  
we find that there exists $\delta_0>0$ such that  
$$ -\epsilon (N_1-1)\frac{(T-{\bar T_1})^2}{T}<-L_1$$ 
for all $T<\delta_0$. Similarly, inspecting the second 
term in $\dot W$ and using Lemma~\ref{minilemma}, we find that there exists $\delta_1>0$ such that  
$$ N_1\beta_1{\bar T_1^*} 
\left[3(1+q_{11}\mu)^{1/3}- 
(1+q_{11}\mu)\frac{V_1 T{\bar T_1^*}} 
{{\bar V_1}{\bar T_1}T_1^*}-\frac{T_1^*{\bar V_1}}{{\bar T_1^*}V_1}- 
\frac{{\bar T_1}}{T}\right]<-L_1$$ 
for all $\frac{T^*_1}{V_1}<\delta_1$ and all $\mu \in [0,\mu_1]$. Finally, 
there exists $\delta_2>0$ such that  
$-\frac{\alpha\eta}{2}+ \alpha V_1 <-\frac{\alpha \eta}{4}$ 
for all $V_1<\delta_2$. 
Let  
$$\hat K_{\delta}=\{ (T,T^*,V) \in K \cap U | T \geq \delta_0, V_1 \geq \delta_2, 
T^*_1 \geq \delta_1 V_1\}.$$ 
Consider $(T,T^*,V) \in (K \cap U) \backslash \hat K_{\delta}$ 
and let $\mu \in [0,\mu_1]$, then at least one of the following holds: 
\begin{itemize} 
\item $T <\delta_0$, in which case $ \dot W \leq -L_1 -\frac{\alpha \eta}{2} + L_1 \leq  
-\frac{\alpha \eta}{2}$; 
\item $T^*_1/V_1 < \delta_1$, in which case $ \dot W \leq -L_1 -\frac{\alpha \eta}{2} + L_1 \leq  
-\frac{\alpha \eta}{2}$; 
\item $ V_1 < \delta_2$, in which case $\dot W \leq  -\frac{\alpha \eta}{4}$; 
\end{itemize} 
Hence, for all $(T,T^*,V) \in (K \cap U) \backslash \hat K_{\delta}$ 
and all $\mu \in [0,\mu_1]$, we have 
$ \dot W \leq -\frac{\alpha \eta}{4}<0.$ From this point forward, the proof is  
identical to the proof of Theorem~\ref{compactset}, so it will be omitted. 
 
\noindent (b) Now we consider system $(\ref{hiv3b})$. Let $W$ be the same as in part (a). 
Computing $\dot W$ for the system $(\ref{hiv3b})$, we obtain 
\begin{eqnarray*} 
 \dot W &=& (N_1-1)\left[f'(c)+\frac{k}{\gamma}f({\bar T_1}) \right]\frac{(T-{\bar T_1})^2}{T} 
+N_1\beta_1{\bar T_1^*}\left[3-\frac{V_1 T{\bar T_1^*}} 
{{\bar V_1}{\bar T_1}T_1^*}-\frac{T_1^*{\bar V_1}}{{\bar T_1^*}V_1}-\frac{{\bar T_1}}{T}\right]\\ 
 & & -(N_1-1) \sum_{i=2}^n k_i V_i (\bar T_i -\bar T_1) +  
\mu \frac{V_1-\bar V_1}{V_1} \sum_{j=1}^n q_{1j} N_j \beta_j T^*_j 
+\mu \sum_{i=2}^n \frac{N_1-1}{N_i-1} \sum_{j=1}^n q_{ij} N_j \beta_j T^*_j, 
\end{eqnarray*} 
Recombining the terms, we find that 
\begin{eqnarray*} 
\dot W &=& (N_1-1)\left[f'(c)+\frac{k}{\gamma}f({\bar T_1}) \right]\frac{(T-{\bar T_1})^2}{T} 
+N_1\beta_1{\bar T_1^*}\left[3-\frac{V_1 T{\bar T_1^*}} 
{{\bar V_1}{\bar T_1}T_1^*}-\frac{T_1^*{\bar V_1}}{{\bar T_1^*}V_1}-\frac{{\bar T_1}}{T}\right]\\ 
 & & -(N_1-1) \sum_{i=2}^n k_i V_i (\bar T_i -\bar T_1)  
+\mu \sum_{i=1}^n \frac{N_1-1}{N_i-1} \sum_{j=1}^n q_{ij} N_j \beta_j T^*_j \\ && 
-\mu q_{11} \frac{\bar V_1 N_1 \beta_1 T^*_1}{V_1} -  
\mu \frac{\bar V_1}{V_1} \sum_{j=2}^n q_{1j} N_j \beta_j T^*_j, 
\end{eqnarray*} 
where the last term is clearly non-positive. Let  
\begin{eqnarray*} 
 \alpha &= &(N_1-1) \min_{i\geq 2} k_i (\bar T_i -\bar T_1) >0,\\ 
 L & =& \sup_{K} \sum_{i=1}^n \frac{N_1-1}{N_i-1} \sum_{j=1}^n q_{ij} N_j \beta_j T^*_j \geq 0. 
\end{eqnarray*} 
By Lemma~\ref{lowerboundKVT}, 
 there exist $\eta,\mu_a>0$ such that $1'V(t) > \eta$  
for all $\mu \in [0,\mu_a]$ 
 and all sufficiently large $t$. Hence, by shifting 
time forward if necessary, we have the inequality 
\begin{eqnarray*} 
\dot W &\leq & -\epsilon (N_1-1)\frac{(T-{\bar T_1})^2}{T}+N_1\beta_1{\bar T_1^*}\left[3-\frac{V_1 T{\bar T_1^*}} 
{{\bar V_1}{\bar T_1}T_1^*}-\frac{T_1^*{\bar V_1}}{{\bar T_1^*}V_1}-\frac{{\bar T_1}}{T}\right]\\ 
 & & -\alpha (\eta -V_1) +\mu L -\mu q_{11} \frac{\bar V_1 N_1 \beta_1 T^*_1}{V_1}, 
\end{eqnarray*} 
which holds in $K$ for all $\mu \in [0,\mu_a]$. We combine the second and the last 
terms to obtain 
\begin{eqnarray*} 
\dot W &\leq & -\epsilon (N_1-1)\frac{(T-{\bar T_1})^2}{T}+N_1\beta_1{\bar T_1^*}\left[3- 
\frac{V_1 T{\bar T_1^*}} 
{{\bar V_1}{\bar T_1}T_1^*}-(1+q_{11}\mu)\frac{T_1^*{\bar V_1}}{{\bar T_1^*}V_1}- 
\frac{{\bar T_1}}{T}\right]\\ 
 & & -\alpha (\eta -V_1) +\mu L. 
\end{eqnarray*} 
Further, we rewrite the above inequality as 
\begin{eqnarray*} 
\dot W &\leq & -\epsilon (N_1-1)\frac{(T-{\bar T_1})^2}{T}+N_1\beta_1{\bar T_1^*} 
\left[3(1+q_{11}\mu)^{1/3}- 
\frac{V_1 T{\bar T_1^*}} 
{{\bar V_1}{\bar T_1}T_1^*}-(1+q_{11}\mu)\frac{T_1^*{\bar V_1}}{{\bar T_1^*}V_1}- 
\frac{{\bar T_1}}{T}\right]\\ 
 & & -\alpha (\eta -V_1) +\mu L + 3 N_1\beta_1{\bar T_1^*}\left[1-(1+q_{11}\mu)^{1/3}\right]. 
\end{eqnarray*} 
Let $\mu_b>0$ be such that for all $\mu \in [0,\mu_b]$, 
$$ (1+q_{11}\mu) \in [\frac{1}{2},1], \quad 
-\alpha \eta +\mu L + 3 N_1\beta_1{\bar T_1^*}\left[1-(1+q_{11}\mu)^{1/3}\right]  
\leq -\frac{\alpha \eta}{2} .$$ 
Now we let $\mu_1=\min[\mu_a,\mu_b]$, so that for all $\mu \in [0,\mu_1]$ 
and all points in $K$, 
\begin{eqnarray*} 
\dot W &\leq & -\epsilon (N_1-1)\frac{(T-{\bar T_1})^2}{T}+N_1\beta_1{\bar T_1^*} 
\left[3(1+q_{11}\mu)^{1/3}- 
\frac{V_1 T{\bar T_1^*}} 
{{\bar V_1}{\bar T_1}T_1^*}-(1+q_{11}\mu)\frac{T_1^*{\bar V_1}}{{\bar T_1^*}V_1}- 
\frac{{\bar T_1}}{T}\right]\\ 
 & & -\frac{\alpha \eta}{2} +\alpha V_1. 
\end{eqnarray*} 
From this point forward, the proof is  
identical to the proof of part (a), so it will be omitted. 
\end{proof}

\begin{stel}\label{global2} 
 
Let the assumptions of Lemma~\ref{invertibleKVT} hold, let  
$U$ be the set from Theorem $\ref{compactset-KVT}$, and define  
 
$$ U'=\{ (T,T^*,V) \in \reals^{2n+1}_+ | \; \; T^*_1+V_1>0 \}\supset U.$$ 
 
Then there exist $\mu_0>0$ and a continuous map $E: [0,\mu_0] \rightarrow U$  
such that  
\begin{enumerate} 
 
\item 
 $E(0)=E_1$ (where $E_1$ is the same as in Lemma~\ref{invertibleKVT}),  
and  
$E(\mu)$ is an equilibrium of $(\ref{hiv3a})$ or of $(\ref{hiv3b})$ for  
all $\mu \in [0,\mu_0]$; 
\item For each $\mu \in [0,\mu_0]$,  $E(\mu)$ is  
a globally asymptotically stable equilibrium  of $(\ref{hiv3a})$ or  
of $(\ref{hiv3b})$ in $U'$. 
 
\end{enumerate} 
\begin{proof} 
The proof is similar to that of Theorem $\ref{global}$. 
\end{proof} 
 
\end{stel}


\begin{thebibliography}{199}  

\bibitem{hiv} P. De Leenheer, and H.L. Smith, 
Virus dynamics: a global analysis, SIAM J. Appl. Math. 64 (2003), 1313-1327. 
 
\bibitem{hofbauer} J. Hofbauer, and K. Sigmund, 
Evolutionary Games and Replicator Dynamics, Cambridge University Press, 1998. 
 
\bibitem{hutson1} V. Hutson, A theorem on average Liapunov functions,  Mh. Math. 98 (1984), 267-275. 

 \bibitem{hutson2} V. Hutson, and K. Schmitt, Permanence and the dynamics of biological systems,   
Math. Biosc. 111 (1992), 1-71. 

\bibitem{iggidr} A. Iggidr, J.-C. Kamgang, G. Sallet, and J.-J. Tewa, 
Global analysis of new malaria intrahost models with a competitive exclusion principle,  
SIAM J. Appl. Math. 67 (2006), 260-278. 

\bibitem{korobeinikov} A. Korobeinikov, Lyapunov functions and global properties for SEIR and SEIS  
epidemic models,  IMA Math. Med. Biol. 21 (2004), 75-83.

\bibitem{lasalle} J.P. LaSalle, Stability theory for ordinary differential equations,  
J. Diff. Eqns. 4 (1968), 57-65.

\bibitem{li-muldowney} M.Y. Li, and J.S. Muldowney,  
Global stability for the SEIR model in epidemiology, 
Math. Biosc. 125 (1995), 155-164. 

\bibitem{nowak-may} M.A. Nowak, and R.M. May, 
Virus dynamics, Oxford University Press, New York, 2000. 

\bibitem{perelson-etal} A.S. Perelson, D.E. Kirschner, and R. De Boer, 
Dynamics of HIV infection of CD$4^+$ T cells, Math. Biosc. 114 (1993), 81-125. 
 
\bibitem{perelson-nelson} A.S. Perelson, and P.W. Nelson, 
Mathematical analysis of HIV-1 dynamics in vivo, SIAM Rev. 41 (1999), 3-44. 

\bibitem{zhilan} L. Rong, Z. Feng, and A.S. Perelson, 
Emergence of HIV-1 drug resistance during antiretroviral treatment, preprint. 

\bibitem{smith-waltman-chem} H.L. Smith, and P. Waltman, 
The Theory of the Chemostat: Dynamics of Microbial Competition, Cambridge University Press,  1994. 

\bibitem{smith-waltman} H.L. Smith, and P. Waltman, Perturbation of a globally stable steady state,  
Proc. Amer. Math. Soc. 127 (1999), 447-453.  

\bibitem{thieme93}  H.R. Thieme, Persistence under relaxed point-dissipativity (with application 
to an endemic model), SIAM J. Math. Anal. 24 (1993), 407-435. 
 
\bibitem{wang-li} L. Wang, and M.Y. Li, 
Mathematical analysis of the global dynamics of a model for HIV infection of CD$4^+$ T cells, 
Math. Biosc.  200 (2006), 44-57. 
 
\end{thebibliography}
\end{document}